\documentclass[prb,aps,twocolumn,reprint,amsmath,amssymb,showpacs,superscriptaddress]{revtex4-1}

\usepackage{verbatim}
\usepackage{bm}
\usepackage{dcolumn}
\usepackage{graphicx}
\usepackage{epsfig}
\usepackage{amsfonts}
\usepackage{amsmath}
\usepackage{amssymb}
\usepackage{subfigure}

\usepackage{wasysym}

\usepackage{booktabs}
\newcommand{\ra}[1]{\renewcommand{\arraystretch}{#1}} 

\usepackage{color}

\definecolor{darkblue}{rgb}{0,0.02,0.45}

\usepackage[colorlinks,bookmarks=false,citecolor=blue,linkcolor=red,urlcolor=blue]{hyperref}

\newcommand{\be}{\begin{equation}}
\newcommand{\ee}{\end{equation}}
\newcommand{\bea}{\begin{eqnarray}}
\newcommand{\eea}{\end{eqnarray}}
\newcommand{\fig}[1]{Fig.~\ref{#1}} 
\newcolumntype{.}{D{.}{.}{-1}}

\def\vec{\mathbf}
\def\mc{\mathcal}

\allowdisplaybreaks

\begin{document}
\title{Frustrated magnetism and resonating valence bond physics\\ in two-dimensional kagome-like magnets}

\author{Ioannis Rousochatzakis}
\affiliation{Institute for Theoretical Solid State Physics, IFW Dresden, D-01069 Dresden, Germany}
\author{Roderich Moessner}
\affiliation{Max Planck Institut f$\ddot{u}r$ Physik Komplexer Systeme, N\"othnitzer Str. 38, 01187 Dresden, Germany}
\author{Jeroen van den Brink}
\affiliation{Institute for Theoretical Solid State Physics, IFW Dresden, D-01069 Dresden, Germany}
\affiliation{Department of Physics, TU Dresden, D-01062 Dresden, Germany}

\date{October 18, 2013}

\begin{abstract}
We explore the phase diagram and the low-energy physics of three Heisenberg antiferromagnets which, like the kagome lattice, are networks of corner-sharing triangles but contain two sets of inequivalent short-distance resonance loops. We use a combination of exact diagonalization, analytical strong-coupling theories and resonating valence bond approaches, and scan through the ratio of the two inequivalent exchange couplings. In one limit, the lattices effectively become bipartite, while at the opposite limit heavily frustrated nets emerge. In between, competing tunneling processes result in short-ranged spin correlations, a manifold of low-lying singlets (which can be understood as localized bound states of magnetic excitations), and the stabilization of valence bond crystals with resonating building blocks.  
\end{abstract}

\maketitle

\section{Introduction}
\begin{figure*}[!t]
\includegraphics[width=0.8\textwidth]{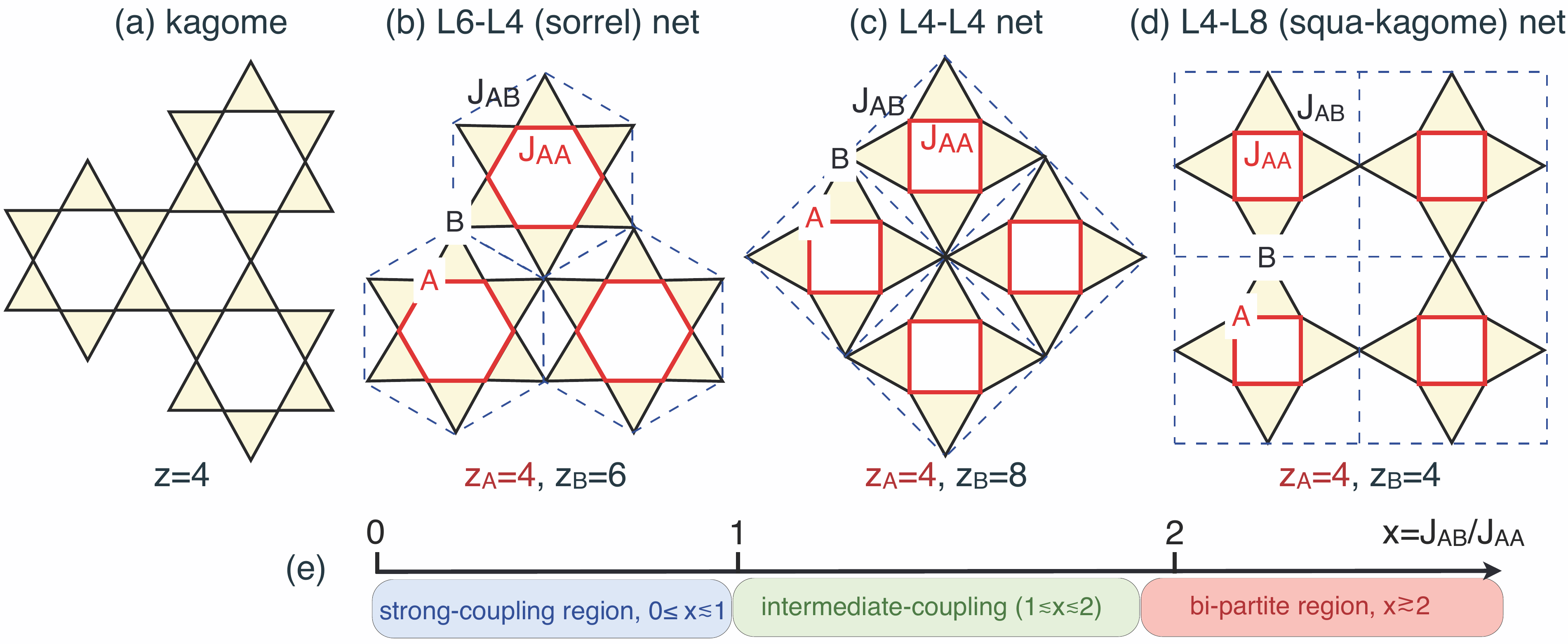} 
\caption{(color online) (a) Kagome lattice. (b)-(d) Corner-sharing-triangle lattices with two types of sites (A and B), two exchange couplings ($J_{\sf AA}$ and $J_{\sf AB}$) and two types of elementary loops. (e) The three generic regimes in $x\!=\!J_{\sf AB}/J_{\sf AA}$.}
\label{fig:Lattices}
\end{figure*}

Highly frustrated antiferromagnets (AFMs) realize exotic collective states of matter in both experiment~\cite{Han12,Mendels07,Clark13,Okamoto07,Balents10} and theory, ranging from valence bond crystals~\cite{MarstonZeng91,NikolicSenthil03,SinghHuse07,Mambrini2010ab,Syromyatnikov,Budnik,Vidal2010,PoilblancMisguich2011} to chiral~\cite{Messio2012,Capponi2012} and Z$_2$ topological spin liquids carrying fractionalized excitations~\cite{YanHuseWhite,Shollwock,Misguich02,Misguich03,Lu,Iqbal,Balents10}. A long-established and very powerful theoretical starting point to capture such magnetically disordered quantum phases is the short-range resonating valence bond basis~\cite{Anderson73, FazekasAnderson74,Anderson87, Sutherland,LiangDoucotAnderson88,Sandvik05}. Its physical relevance is  particularly clear in spin $S\!=\!1/2$ Heisenberg AFMs that are built from triangular units. Pairing any two spins of a triangle into a valence bond (VB) -- a quantum-mechanical singlet -- minimizes the local energy. An essential point is that on lattices with corner-sharing triangles, such as the kagome, the VB covering is geometrically frustrated: a finite fraction of triangles, the {\it defect triangles}, must do without a singlet~\cite{Elser}. This  frustration is due to the presence of closed loops, for instance the loop around a hexagon of the kagome lattice, cf. \fig{fig:Lattices}(a). The ensuing defect triangles are the source of non-trivial VB dynamics. They cause the coherent evolution from one dimer covering into another while the total number of VBs present on the lattice stays the same. 

As initially shown by Zeng and Elser~\cite{ZengElser95}, the low-energy VB dynamics of the $S\!=\!1/2$ kagome magnet is governed by tunneling events of defect triangles across the hexagon loops~\cite{MambriniMila2000,Misguich02,Mambrini2010ab}, while the actual ground state is extremely sensitive to the competition between different such tunneling processes~\cite{Mambrini2010ab}. This calls for a deeper study of the kind of exotic phases which arise from controlling and tuning the most relevant local tunneling amplitudes, while preserving the basic two-dimensional (2D) corner-sharing-triangle layout of the lattice. 

Here we consider the simplest generalizations to kagome-type lattices that naturally accommodate loops of different lengths, which in turn introduce another degree of tunneling competition. The three 2D triangle-based lattices that we consider are shown in Fig.~\ref{fig:Lattices}(b-d). Each of the lattices features two sets of inequivalent sites, labelled A and B. On all lattices the A-sites  are four-fold coordinated ($z_{\sf A}\!=\!4$) as sites on the kagome, while the B-sites have a coordination $z_{\sf B}\!\geq\! 4$. The inequivalent sites cause the presence of two types of short resonance loops: whereas the kagome lattice has one minimal loop of length six (denoted as L6 from here on), we now have loop lengths of L6-L4 ({\it sorrel} net~\cite{HopkinsonBeck}), L4-L4 and L4-L8 ({\it squa-kagome} net~\cite{SquaKagomelargeN,SquaKagomeRichter1,SquaKagomeRichter2,SquaKagomeRichter3,SquaKagomeRichter4}). The resulting $S\!=\!1/2$ AFM Heisenberg Hamiltonian 
\be
\mc{H}\!=\!\sum_{\langle ij\rangle} J_{ij} \vec{S}_i\!\cdot\!\vec{S}_j~,
\ee
where $\langle ij\rangle$ denotes nearest neighbor (NN) spins, is characterized by two inequivalent NN exchange parameters $J_{\sf AA}$ and $J_{\sf AB}$, their ratio $x\!=\!J_{\sf AB}/J_{\sf AA}$ setting the ``coupling strength'' of the model. We study the phase diagrams of these Hamiltonians using a combination of exact diagonalization (ED), analytical strong-coupling theories, and resonating valence bond (RVB) approaches based on quantum dimer model (QDM) derivations. Away from the intermediate coupling regime $1\!\lesssim\! x\!\lesssim\!2$, we find a number of generic features, including Lieb ferrimagnetism, fractional magnetization plateaux, exactly and almost exactly localized modes, as well as physics governed by an effectively frustrating NN and next-NN exchange ($J_1$ and $J_2$, respectively), with $J_2\!\gg\!J_1$. 

At intermediate coupling, strong frustration leads to short-ranged spin correlations and a manifold of low-lying singlets, most of which correspond to localized bound states of triplets or quintets. For the L6-L4 net, the ED results reveal the presence of a valence bond crystal (VBC) state which breaks the rotational symmetry, in agreement with the prediction from the corresponding QDM. In the two other lattices we find evidence for a number of intermediate phases whose exact nature remains unclear. The behavior in the different coupling regimes are presented in detail in turn below. Technical details and supplementing material, including a discussion of the classical ground state manifold, are relegated to the appendices.

\section{Bi-partite regime}
\begin{figure*}
\includegraphics[width=0.99\textwidth]{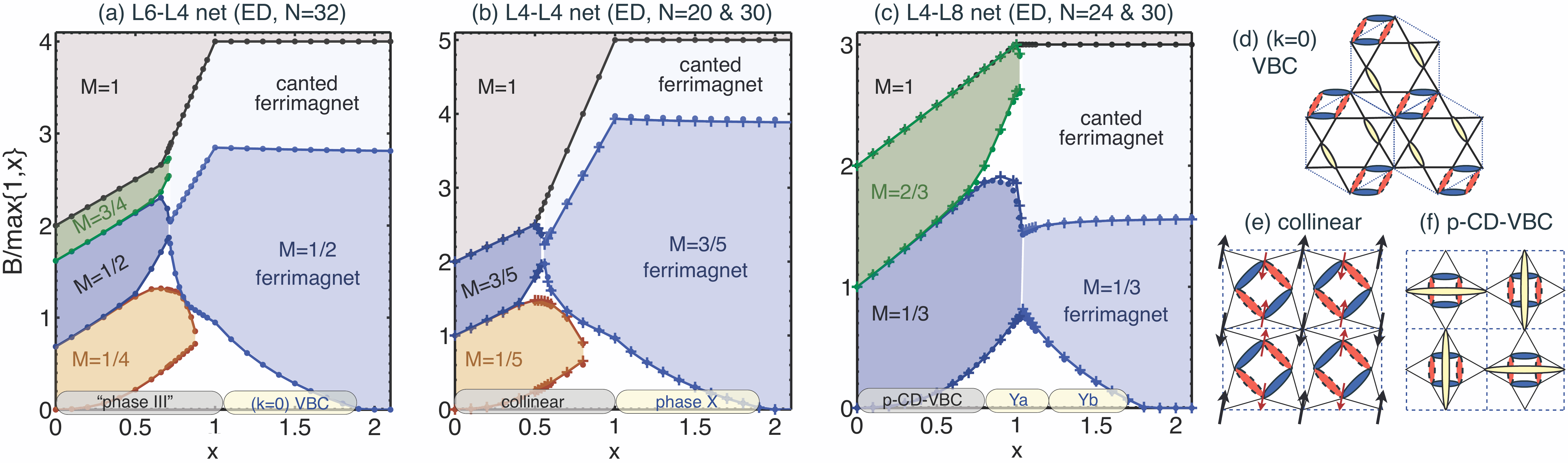}
\caption{(color online) (a-c) $x$-B phase diagram of the three Heisenberg magnets of this study. (b-c) ED results from two cluster sizes, with crosses (dots) corresponding to the largest (smallest) cluster. Ferrimagnetic and corresponding plateaux phases with the same moment are separated by first-order transition lines. (d-f) Schematic representations of states in (a-c). Solid (blue)-dashed (red) ovals denote the symmetric perfect square resonance state $|{\sf PR}_+\rangle$ with $S_{\sf AA}\!=\!0$. Yellow ovals denote VBs.}\label{fig:PhaseDiagram}
\end{figure*}

For $x\!\gg\!1$ all three lattices become effectively bi-partite, leading to semiclassical collinear N\'eel states with, say, the A-spins pointing up and the B-spins down. Since $N_{\sf A}\!\neq\!N_{\sf B}$, these are Lieb ferrimagnets~\cite{Lieb} with total moments M=1/2, 3/5, and 1/3 for L6-L4, L4-L4, and L4-L8 respectively. Classically the onset of this phase is $x_\text{ferri}\!=\!2$ (see App.~\ref{app:clas}), while our ED results show a slightly smaller $x_\text{ferri}$ in all lattices, reflecting a general preference of quantum fluctuations for collinear spin arrangements. In addition, in presence of a magnetic field $B$, these phases survive in a large $x$-$B$ region of the phase diagram, see Fig.~\ref{fig:PhaseDiagram}.  Quantum fluctuations reduce the local spin lengths $S_{\sf A}$ and $S_{\sf B}$, but in such a way that the total moment $M$ is conserved. This means that the deviation $\delta S$ of the spin lengths from their classical $S\!=\!1/2$ value satisfies the ``conservation law''  $\delta S_{\sf B}\!/\!\delta S_{\sf A}\!=\!N_{\sf A}\!/\!N_{\sf B}$~\cite{footnote1}. Our ED results give $(S_{\sf A}, S_{\sf B})\!\simeq\!(0.4545, 0.3635)$, $(0.472049,0.388196)$, and $(0.4084,0.3168)$ for L6-L4 (32 sites), L4-L4 (20 sites), and L4-L8 (30 sites) respectively.

\section{Strong-coupling regime} 
\subsection{Effective models for the ground state}\label{sec:EffTheories} 
The opposite limit, $x\!\ll\!1$, is much richer. Here the A-spins form isolated compound objects (hexamers for L6-L4, tetramers for L4-L4 and L4-L8) with a total spin $S_{\sf AA}\!=\!0$, while the B-spins are free to point up or down, defining a highly degenerate manifold. An infinitesimal $J_{\sf AB}$ mediates an effective coupling between the B-spins through the virtual excitations of the A-spins out of their $S_{\sf AA}\!=\!0$ state.  Degenerate perturbation theory shows that the leading exchange between the B-spins is not a NN exchange $J_1$ but a next-NN coupling $J_2$, which in conjunction with the underlying topology of the B-spins, gives rise to rich frustration effects. This stems from a destructive interference mechanism which is related to the resonating VB nature of the $S_{\sf AA}\!=\!0$ objects, and which cancels out exactly (L4-L4 and L4-L8) or almost exactly (L6-L4) the leading contribution to $J_1$. A similar mechanism appears in the strong dimer limit of the Cairo pentagonal AFM~\cite{pentag1}. For L6-L4, we obtain a $J_2$-$J_3$-$J_1$ model on the honeycomb lattice~\cite{Fouet,Albuquerque11, Cabra2011,LiBishopRichter12} with 
\bea
J_2 \!&=&\! 0.115037 x^2 + 0.10576478 x^3+\mc{O}(x)^4,  \nonumber\\
J_3 \!&=&\! 0.0384319 x^2 + 0.030712121 x^3+\mc{O}(x)^4, \\
J_1 \!&=&\! 2 (-0.019216 x^2 + 0.0553166527 x^3 )+\mc{O}(x)^4 ~.\nonumber 
\eea
The relative strengths of these couplings places the GS in the spiral ``phase-III'' region of Refs.~[\onlinecite{Fouet}-\onlinecite{Albuquerque11}]. 

The L4-L4 and L4-L8 nets share the same $S_{\sf AA}\!=\!0$ plaquette, and so we expect very similar effective couplings. Indeed, an expansion up to third order in $x$ gives an effective $J_2$-$J_1$ model on the square (L4-L4) or the checkerboard (L4-L8) lattice with 
\be\label{eq:J2}
J_2\!=\!\frac{1}{6}x^2+\frac{1}{8}x^3+\mc{O}(x^4),~~
J_1\!=\!f\frac{1}{24}x^3+\mc{O}(x^4)~,
\ee
where $f=2$ (L4-L4) or $1$ (L4-L8). For the L4-L4 net, the finite $J_1$ appearing in third order stabilizes, via a quantum order-by-disorder effect~\cite{ChandraDoucot}, the collinear striped phase of Fig.~\ref{fig:PhaseDiagram}(e) among the one-parameter family of states favored by $J_2$ alone. In L4-L8, on the other hand, the finite $J_1$ stabilizes a crossed-dimer VBC state~\cite{Starykh05,ArlegoBrenig0709,BishopRichter12} which is denoted by p-CD-VBC in \fig{fig:PhaseDiagram}, where p stands for the plaquette structure of the A-spins. This state has a finite spin gap and thus corresponds to an $M\!=\!0$ plateau.

{\it Four-body terms in L4-L4 and L4-L8} --- 
Pushing the expansion up to fourth-order in $x$ gives a four-spin exchange around each AA-square of the form 
\bea
\mc{H}_{\sf K}\!&=&\!K_{\sf h}\parbox{0.3in}{\epsfig{file=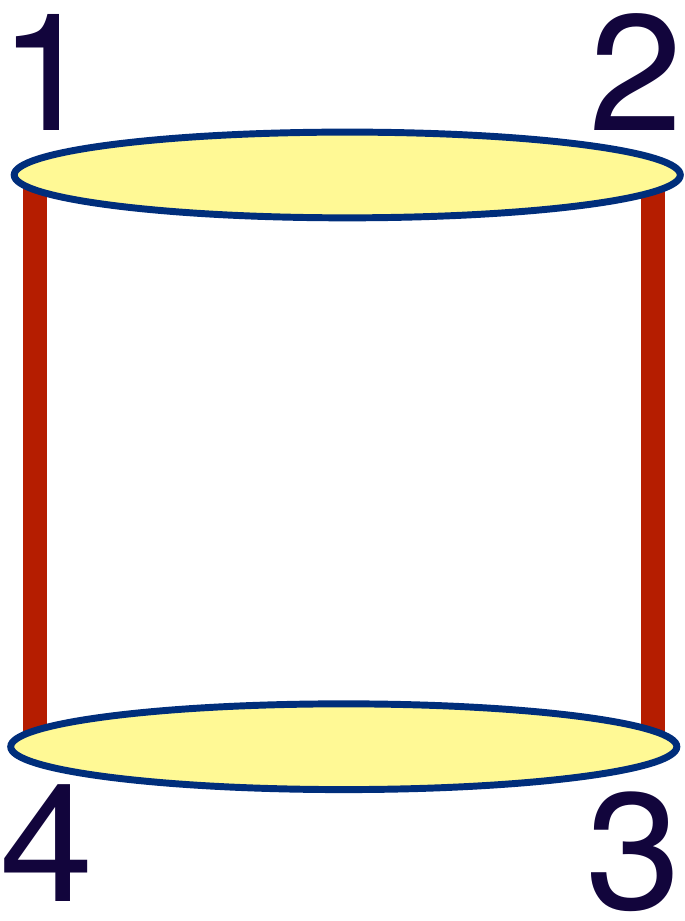,width=0.3in,clip=}}
\!+\!K_{\sf v}\parbox{0.3in}{\epsfig{file=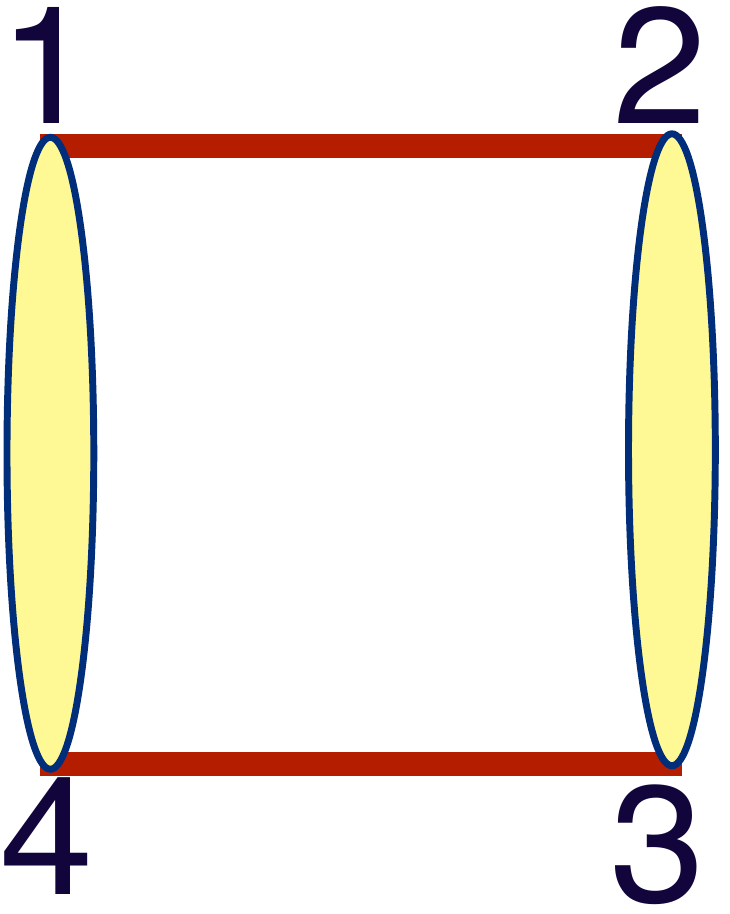,width=0.3in,clip=}}
\!+\!K_{\sf x}\parbox{0.27in}{\epsfig{file=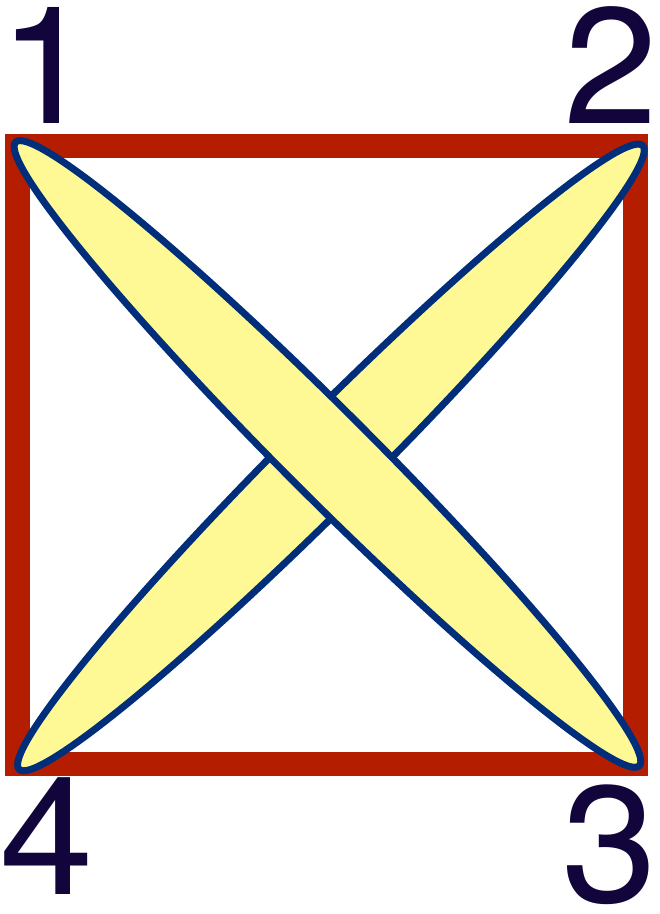,width=0.27in,clip=}}
\!=\!K_{\sf h} (\vec{S}_1\!\cdot\!\vec{S}_2)(\vec{S}_3\!\cdot\!\vec{S}_4)
\nonumber\\
&&\!\!+\!K_{\sf v} (\vec{S}_1\!\cdot\!\vec{S}_4)(\vec{S}_2\!\cdot\!\vec{S}_3)
\!+\!K_{\sf x} (\vec{S}_1\!\cdot\!\vec{S}_3)(\vec{S}_2\!\cdot\!\vec{S}_4)~,
\eea
with $K_{\sf h}\!=\!K_{\sf v}\!=\!-\frac{7}{144} x^4+\mc{O}(x^5)$ and $K_{\sf x}\!=\!+\frac{37}{216} x^4+\mc{O}(x^5)$. Such ring exchange~\cite{RingExchange} type of terms emerge also in the strong dimer limit of the Cairo pentagonal AFM~\cite{pentag1}. 

Let us now discuss the effect of $\mc{H}_{\sf K}$ in the small-$x$ region of L4-L4. Taking the one-parameter family of classical states $|\phi\rangle$ favored by $J_2$ alone~\cite{ChandraDoucot}, where $\phi$ is the angle between the two N\'eel sublattices, we find $\langle\phi|\mc{H}_{K}|\phi\rangle\!=\!(K_{\sf h}\!+\!K_{\sf v})\cos^2\phi\!+\!K_{\sf x}$. In contrast to the Cairo AFM~\cite{pentag1}, here both $K_{\sf h}$ and $K_{\sf v}$ are negative and so $\mc{H}_{K}$ does not compete with $J_1$, i.e. the collinear phase survives as long as $J_2$ remains the dominant term in the effective theory.  

Turning to L4-L8, to see the effect of $\mc{H}_{K}$ we consider a single AA-square and keep only $K_{\sf x}$, as $K_{\sf x}/K_{\sf h,v}\!=\!3.52+\mc{O}(x)$. This term favors two triplet ground states, namely $|T,m\rangle\!=\!|s\rangle_{13}\otimes|t_m\rangle_{24}$ and $|T',m\rangle\!=\!|t_m\rangle_{13}\otimes|s\rangle_{24}$ ($m=0,\pm1$), where $|s\rangle_{ij}$ and $|t_m\rangle_{ij}$ denote the singlet and the three components of the triplet on the bond $(ij)$. This means that $K_{\sf x}$ disfavors having two singlets (or two triplets) on the two diagonals of a plaquette. Such anti-correlation of singlets is already partly present in the p-CD-VBC state of Fig.~\ref{fig:PhaseDiagram}(f), since a diagonal bond without a singlet has a large amplitude (3/4, without quantum fluctuations) in the triplet sector. This suggests that (i) an infinitesimal $K_{\sf x}$ does not frustrate the p-CD-VBC state, and (ii) it enhances the triplet amplitude on empty diagonals~\cite{footnote2}.

{\it Feedback on the A-sites} --- 
According to the above, both in L6-L4 and L4-L4 the B-sites order magnetically at small $x$. It is natural to expect that the A-sites will eventually also attain a small moment owing to the finite exchange fields exerted from the B-sites. These fields are non-uniform and so they immediately admix finite magnetic corrections to the $S_{\sf AA}\!=\!0$ singlets. In L4-L4, for example, half of the A-sites become polarized to linear order in $x$, see thin (red) arrows in Fig.~\ref{fig:PhaseDiagram}(e). This asymmetry is inherited from the spontaneous symmetry breaking of the B-spins. 

\begin{figure*}[!t]
\includegraphics[width=\textwidth]{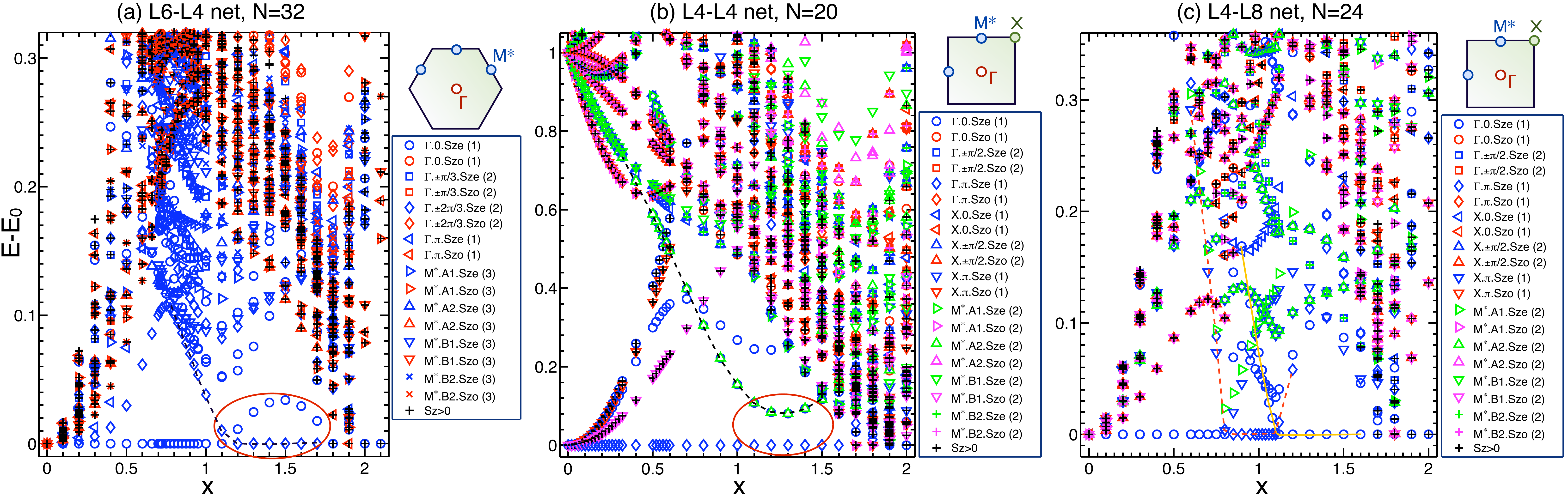} 
\caption{(color online) Low-energy spectra (in units of $J_{\sf AA}$) as a function of $x$ for the (a) 32-site L6-L4 , (b) 20-site L4-L4 (b), and (c) 24-site L4-L8 with periodic boundary conditions. States are labeled by linear momenta, point group quantum numbers, as well as spin inversion symmetry (``Sze'' for even, ``Szo'' for odd) in the $S_z\!=\!0$ sector. Singlets are denoted by open symbols, and magnetic states by symbols with (black) crosses. Ovals in (a) and (b) indicate the group of singlets that form the corresponding intermediate phases in the thermodynamic limit~\cite{footnote55}. Lines are guides to the eye.
}\label{fig:spectra}
\end{figure*} 

\subsection{Field-induced phases \& localized excitations} 
We now consider the systems at strong coupling in the presence of a magnetic field $B$. Given the perturbative scale of their mutual effective couplings, the B-spins will be quickly polarized by a small field. For larger fields, these magnets show successive plateaux at the level-crossing fields of the AA-hexamers (for L6-L4 ) or tetramers (for L4-L4 and L4-L8) from $S_{\sf AA}\!=\!0$ up to $L_{\sf AA}/2$ (where $L_{\sf AA}$ is the length of the shortest AA-loop), see \fig{fig:PhaseDiagram}. 

At saturation, there are exact magnetization jumps which are related to the existence of localized magnons around AA-loops with angular momentum $\pi$. Such magnons arise from a destructive interference mechanism which is generic for corner-sharing-triangle lattices~\cite{Schnack2001,Schulenburg_ind_magn, Richter_chapter, Schmidt_ind_magn, SquaKagomeRichter3, SquaKagomeRichter4}. Here, a simple diagonalization in the one-magnon space shows that localized magnons are present for all $x$, but become the lowest excitations above the fully polarized state only below a characteristic value of $x_c\!=\!\frac{2}{3}$ for L6-L4, $x_c\!=\!\frac{1}{2}$ for L4-L4, and $x_c\!=\!1$ for L4-L8, see \fig{fig:PhaseDiagram}. We also note that the length of the magnetization jump at saturation corresponds to exciting one localized magnon in each AA-loop. This occurs with no extra energy cost since the AA-loops are disconnected.

Besides the exact jumps at saturation, Fig.~\ref{fig:PhaseDiagram} shows almost exact jumps for the transitions between the remaining plateaux. This shows that the corresponding excitations have very small tunneling amplitudes, which can be explained by the fact that they involve large quantum-mechanical objects (an AA-hexamer or AA-tetramer). We also observe that the boundaries between the different plateaux scale linearly with $x$ with slope one. The origin of this behavior is explained in App.~\ref{app:PlateauBoundaries}.

\begin{table*}[!t]
\begin{ruledtabular}
\begin{tabular}{@{}r|cccccccc@{}} 
lattice & $(\frac{N_{\sf A}}{N}, \frac{N_{\sf B}}{N})$ & $(\frac{N_{\sf AA}}{N}, \frac{N_{\sf AB}}{N})$ & $\frac{N_{\sf dt}}{N_{\sf t}}$   & $\mc{D}_{\sf NNVB}$  & $(\frac{N_{\sf vb}^{\sf AA}}{N_{\sf vb}}, \frac{N_{\sf vb}^{\sf AB}}{N_{\sf vb}})$  & $\frac{E^{(0)}}{J_{\sf AA}N}$ & $(e_{\sf AA}^{(0)}, e_{\sf AB}^{(0)})$ & $(e_{\sf AA}, e_{\sf AB})$   \\
\hline
L6-L4 & $(\frac{3}{4},\frac{1}{4})$  &$(\frac{3}{4},\frac{3}{2})$ &$\frac{1}{3}$ & $2\cdot3^{N/4}$ &  $(\frac{1}{2},\frac{1}{2})$ & $-\frac{3}{4}(1+x)$& $(-\frac{1}{4},-\frac{1}{8})$ &  $(-0.2993, -0.1659)$\\
L4-L4 & $(\frac{4}{5},\frac{1}{5})$ &$(\frac{4}{5},\frac{8}{5})$ &$\frac{3}{8}$ & $2\cdot4^{N/5}$ &  $(\frac{3}{5},\frac{2}{5})$ & $-\frac{3}{40}(3+2x)$ & $(-\frac{9}{32},-\frac{3}{32})$ & $(-0.3829, -0.1252)$\\
L4-L8 & $(\frac{2}{3},\frac{1}{3})$ &$(\frac{2}{3},\frac{4}{3})$ &$\frac{1}{4}$ & $2\cdot2^{N/3}$ &  $(\frac{1}{3},\frac{2}{3})$ & $-\frac{3}{24}(1+2x)$ & $(-\frac{9}{32},-\frac{9}{64})$ & $(-0.2556,-0.2084)$\\
\hline
kagome & 1 &  2 & $\frac{1}{4}$  & $2\cdot2^{N/3}$ & 1  & $-\frac{3}{8}$ &  $-\frac{3}{16}$ & $-0.2191(8)$ (ED - 36 sites\cite{Waldtmann}) 
\end{tabular}
\end{ruledtabular}
\caption{Basic elements of the NNVB basis. $N_{\sf A}$ ($N_{\sf B})$: number of type-A (B) sites; $N_{\sf dt}/N_{\sf t}$: number of defect triangles over total number of triangles; $\mc{D}_{\sf NNVB}$: Pauling estimate for the NNVB basis dimension, see App.~\ref{app:nnvb}; $N_{\sf vb}^{\sf AA}$ ($N_{\sf vb}^{\sf AB}$): number of VBs residing on AA (AB) bonds; $N_{\sf vb}\!=\!N/2$: total number of VBs; $e_{\sf AA}^{(0)}$ ($e_{\sf AB}^{(0)}$): average $\langle\vec{S}_i\!\cdot\!\vec{S}_j\rangle$ on AA (AB) bonds over the NNVB states without resonances; $e_{\sf AA}$ ($e_{\sf AB}$): Exact GS expectation values in the 20-site L4-L4 and 24-site L4-L8 clusters at $x\!=\!1$.}
\label{tab:nnvb}
\end{table*}

\subsection{Localized singlets}\label{sec:LocalSinglets}
It turns out that the localized nature of the magnetic excitations has a profound impact on the non-magnetic sector as well. Indeed, one may form an extensive number of singlets by binding two magnetic excitations, one residing on a AA-plaquette and the other on its surrounding BB-plaquette (hexagon for L6-L4, square for L4-L4 and L4-L8). The AA-excitation costs a lot of energy at $x\!=\!0$ but, as we show below, the binding energy scales linearly with $x$ (to leading order), resulting in a high density of low-energy singlets around $x\!\sim\!1$. 

To demonstrate the above we consider the cluster 
\be\label{eq:AAsquare}
\parbox{0.75in}{\epsfig{file=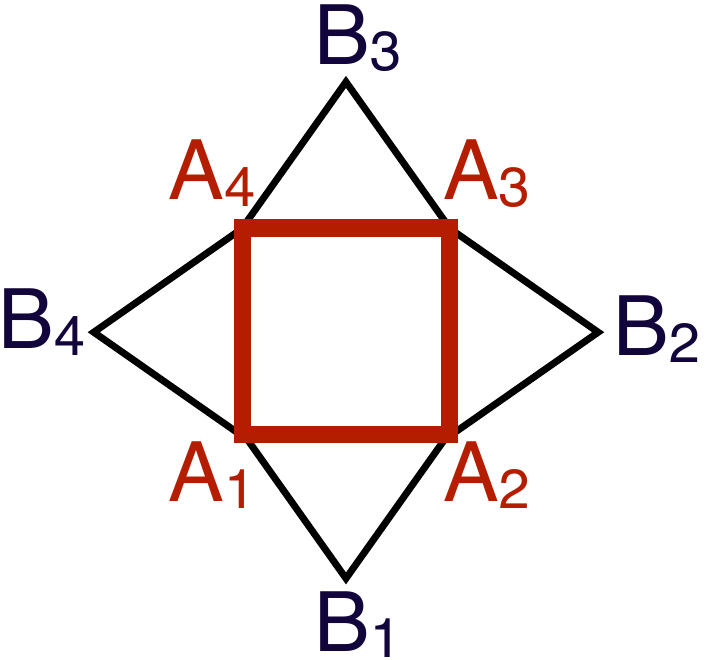,width=0.75in,clip=}}
\ee
which is the basic building block of  L4-L4 or L4-L8 (a similar analysis can be carried out in L6-L4), 
and focus on the manifold of states where the AA-square has been excited to its lowest triplet. In what follows $\vec{A}\!=\!\sum_i\!\vec{A}_i\!=\!1$, $\vec{B}\!=\!\sum_i\!\vec{B}_i$, $\vec{B}_{ij}\!=\!\vec{B}_i\!+\!\vec{B}_j$, and we also introduce the quadrupolar tensor operators $Q_A^{\alpha\beta}$ and $Q_{B_{ij}}^{\alpha\beta}$, see  App.~\ref{app:pert}. To linear order in $x$ we may replace $\vec{A}_i\!\mapsto\!\vec{A}/4$, leading to an effective exchange $\frac{x}{2}\vec{A}\!\cdot\!\vec{B}$, which is the leading term in the binding mechanism. Including quadratic terms (see App.~\ref{app:pert}) gives an effective exchange between B-spins but also quadrupolar couplings between $\vec{A}$ and pairs of B-spins~\cite{footnote3a}: 
\bea\label{eq:AA1}
\mc{H}_{\sf eff}^{(\vec{A}=1)}\!&=&\!c\!+\!J_{\sf AB}\vec{A}\!\cdot\!\vec{B}
\!+\!J_{\sf BB}(\vec{B}_1\!\cdot\vec{B}_3\!+\!\vec{B}_2\!\cdot\vec{B}_4)\nonumber\\
&\!+\!&J_{\sf Q}\left(Q_A^{\alpha\beta}Q_{B_{13}}^{\alpha\beta}\!+\!Q_A^{\alpha\beta}Q_{B_{24}}^{\alpha\beta}\right)
\!+\!\mc{O}(x^3)~,
\eea
where $\alpha,\beta$ are summed over $x,y,z$, $c\!=\!1\!-\!x^2/2$, $J_{\sf AB}\!=\!x/2\!+\!x^2/8$, $J_{\sf BB}\!=\!x^2/3$, $J_{\sf Q}\!=\!x^2/32$, and  we measure energies from the $x\!=\!0$ ground state. 
Diagonalizing (\ref{eq:AA1}) gives three low-energy singlets corresponding to $({\sf B}_{13},{\sf B}_{24}, {\sf B})\!=\!(1,1,1)$, $(1,0,1)$ and $(0,1,1)$, with $E_{1,2}\!=\!c\!-\!2J_{\sf AB}\!\pm\!J_{\sf BB}/2\!\mp\!20J_{\sf Q}/3$ and $E_{3}=E_2$. So for all  singlets the binding energy scales as $1\!-\!x$ at small $x$, as announced above. 

Now, since the AA-triplets are almost localized the same is expected for the singlets, meaning that there is an extensive number of them. The situation is actually much richer, as we may take $n$ localized bound states residing around different~\cite{footnote3b} AA-squares. To linear order in $x$ the energy of such states is $E_n\!=\!n E_1$, and so they should all have low energies at intermediate $x$. 

In a similar fashion, there also exist bound states between an AA-quintet and a BB-quintet. Here the effective interaction reads $\mc{H}_{\sf eff}^{({\sf A}=2)} = 3+ \frac{x}{2}\vec{A}\!\cdot\!\vec{B}+\mc{O}(x^2)$. The lowest singlet corresponds to ${\sf B}\!=\!2$ with $E_1'\!=\!3(1-x)$, and again we may excite $n$ such bound states with $E_n'\!=\!n E_1'$. 

Due to their localized nature, it is natural to think of the above singlets as resonances between short-range VB states. In this sense, the binding mechanism ``pre-forms'' the VB states that are expected to appear around $x\!\sim\!1$, owing to the corner sharing triangle topology. To make this connection more explicit we remark that the above singlet 
with $({\sf B}_{13}, {\sf B}_{24}, {\sf B})\!=\!(1,1,1)$ has a very large overlap with the so-called pinwheel state $|{\sf PW}_-\rangle$ (see below and Eq.~(\ref{eq:PWpm}) of App.~\ref{app:nnvb}), while the $(1,0,1)$ or $(0,1,1)$ singlets are the 6-loop resonances $|{\sf L}6_+\rangle$ of (\ref{eq:L6pm}) along the vertical or the horizontal direction of (\ref{eq:AAsquare}), respectively.

\begin{figure}[!b]
\includegraphics[width=0.45\textwidth]{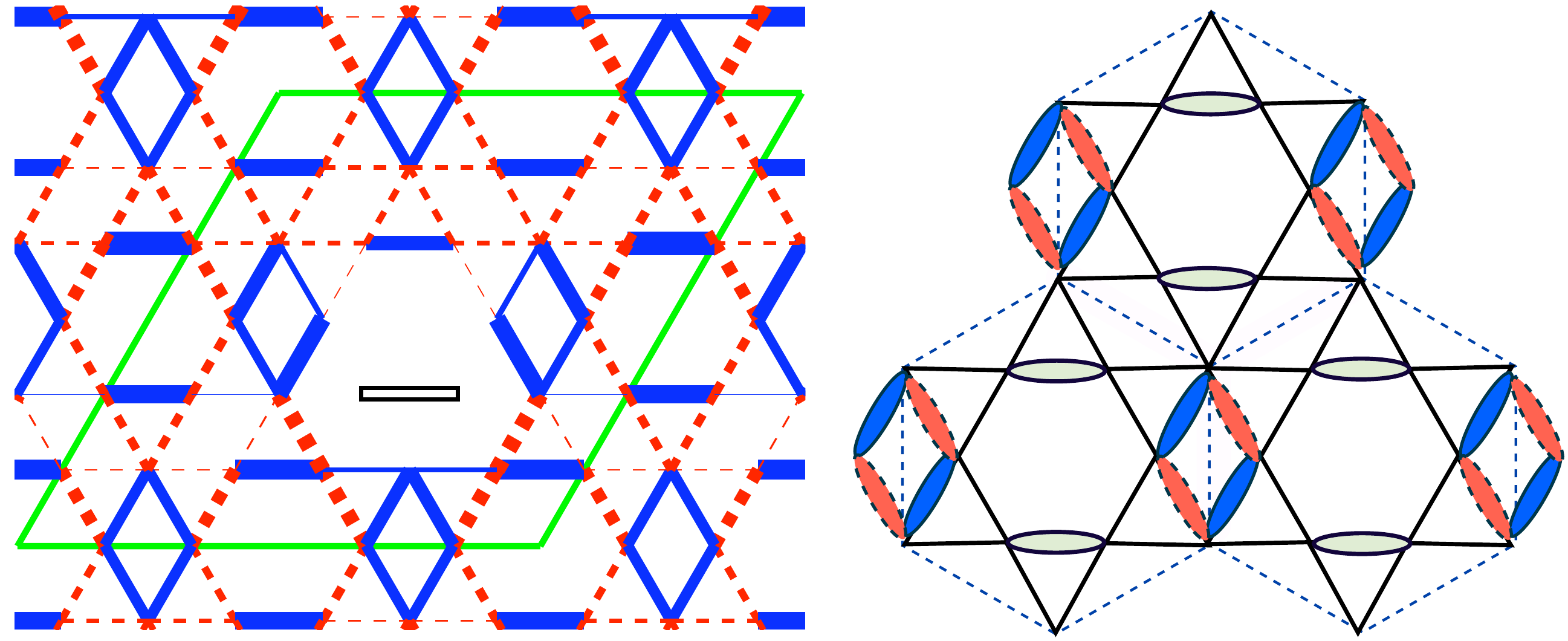} 
\caption{(color online) Left: GS expectation values of the connected dimer-dimer correlations for the 32-site L6-L4 cluster (enclosed by the green parallelogram) at $x\!=\!1.2$. The reference bond is denoted by the thin rectangle. Positive (negative) values are denoted by solid blue (dashed red) lines, and the magnitude scales with the width of the lines. Right: Same as in Fig.~\ref{fig:PhaseDiagram}(d) for comparison.}\label{fig:LRDimer}
\end{figure} 

\section{Intermediate coupling}
This is the most challenging regime. As in the kagome~\cite{Lecheminant97,Waldtmann,Sindzingre2000}, spin-spin correlations are short-ranged and we also observe an accumulation of many low-lying singlets below the lowest magnetic excitations, see Figs.~\ref{fig:spectra} and \ref{fig:spectra30sites}. Specifically, at $x\!=\!1$ we find: 49 singlets for the 32-site L6-L4, 5 (resp. 15) singlets for the 20-site (resp. 30-site) L4-L4, and 18 (resp. 27) singlets for the 24-site (resp. 30-site) L4-L8 net~\cite{footnote50}. As discussed above, many of the low-lying singlets can be understood as localized bound states of two magnetic excitations. A quick inspection of the exact spectra as a function of $x$ shows indeed that some of the low-lying singlets around $x\!\sim\!1$ are connected to the $E\!=\!1$ manifold of $x\!=\!0$ with $E\!=\!1-x+\mc{O}(x^2)$. Other singlets are connected to higher manifolds, as discussed above. 

Now, one or more of these singlets approach quickly and remain close to (or even cross) the GS above $x\!\sim\!1$, which is evidence for phase transitions. Examining the low-lying states and the GS level-crossings in Figs.~\ref{fig:spectra} and \ref{fig:spectra30sites} suggests one intermediate phase in L6-L4, one in L4-L4, and at least two in L4-L8. In Fig.~\ref{fig:PhaseDiagram} these states are denoted by ``(k=0)\! VBC'', ``X'', and ``Ya''-``Yb'' respectively. 

To further probe the nature of these phases we now turn to the nearest neighbor VB (NNVB) description. The general features of the NNVB basis are provided in Table~\ref{tab:nnvb}. We first check that the asymmetric distribution of the GS energy on the two types of bonds agrees qualitatively with the corresponding distribution over the NNVB basis states without resonances, see last two columns of Table~\ref{tab:nnvb}. For the particular case of L4-L4, the exact results for the energy distribution show clearly that singlets prefer to reside on AA-bonds rather than on AB-bonds due to the high connectivity of the B-sites.

\begin{figure}[!b]
\includegraphics[width=0.4\textwidth]{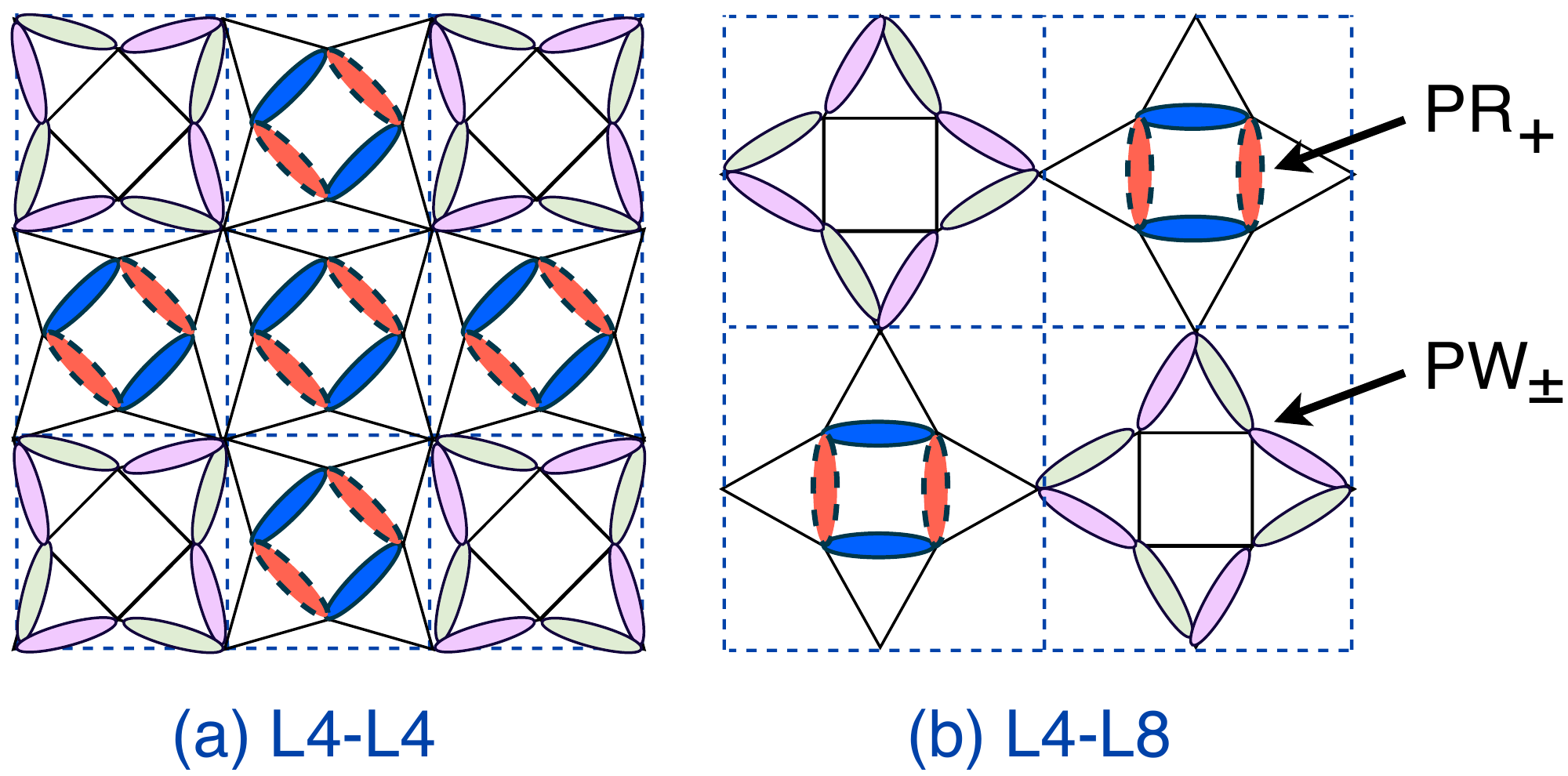} 
\caption{VBC states with the maximum number of ``perfect AA-square resonances'' $|{\sf PR}_+\rangle$ on AA-squares in L4-L4 and L4-L8. The remaining, octagonal dimerized loops stand for the ``AA-pinwheels'' $|{\sf PW}_{\pm}\rangle$ of (\ref{eq:PWpm}).}\label{fig:pinwheels}
\end{figure}

To include the effect of resonances we proceed in analogy with the kagome~\cite{ZengElser95,MambriniMila2000, Mambrini2010ab} and cast the problem in terms of tunneling events of defect triangles across the shortest loops. The resonances around a hexagon at $x\!=\!1$ are known from the kagome~\cite{Mambrini2010ab}, but need to be re-evaluated for $x\!\neq\!1$. The processes around a square involve tunneling of zero, one, or two defect triangles, while on an octagon they involve up to four defect triangles. We have evaluated all resonance amplitudes using a method that essentially corresponds to the infinite order overlap expansion of [\onlinecite{Mambrini2010ab}]. The numerical values for the most local processes are provided in Table~\ref{tab:qdm} of App.~\ref{app:nnvb}, and give, in conjunction with ED results, the following insights for our lattices. 

\subsection{L6-L4 net}
Here the strongest tunneling amplitude around AA-hexagons is $t^{\sf AA}_6\!=\!\frac{3}{5}$ while the one around AB-squares is $t^{\sf AB}_4\!=\!-x$. So, at intermediate $x$ and above, the system will try to maximize the number of AB-square resonances.  This is achieved by the translationally invariant VBC state of \fig{fig:PhaseDiagram}(d) which breaks the six-fold rotational symmetry down to two-fold, in agreement with the symmetry pattern of the three lowest singlets of \fig{fig:spectra}(a) (highlighted by a red oval). The connected dimer-dimer ground state correlations of \fig{fig:LRDimer} are also fully compatible with the dimerization pattern of Fig.~\ref{fig:PhaseDiagram}(d), giving strong confidence that this VBC is indeed stabilized in the thermodynamic limit. 

\subsection{L4-L4 and L4-L8 nets}
Here the dominant tunneling amplitudes and phase space arguments suggest that, in both L4-L4 and L4-L8, the NNVB physics is controlled by the perfect AA-square resonances. Indeed, in L4-L8 we have $t^{\sf AB}_8\!=\!-\frac{8}{21}x$, which is smaller than $t^{\sf AA}_4\!=\!-1$ up to the ferrimagnetic region. On the other hand, in L4-L4 we have two ``perfect square'' processes with $t^{\sf AA}_4\!=\!-1$ and $t^{\sf AB}_4\!=\!-x$. So at first sight one expects that resonances take place mostly around AB-squares for $x\!>\!1$. However, the phase space for such resonances is very small due to the large connectivity of the B-sites. Furthermore, the next-leading 6-loop processes have a very small amplitude $t^{\sf AA}_6=\frac{1}{5}(2x-1)$. 

Maximizing the perfect AA-squares in L4-L4 and L4-L8 leads to the VBC states of Fig.~\ref{fig:pinwheels}. Comparing with the strong coupling limit, here some of the AA-singlets have turned into ``square pinwheels'' $|{\sf PW}_{\pm}\rangle$ (see Eq.~(\ref{eq:PWpm}) of App.~\ref{app:nnvb}), which are analogous to the ``hexagonal pinwheels'' of the 36-site VBC candidate of the kagome~\cite{MarstonZeng91,NikolicSenthil03,SinghHuse07,Mambrini2010ab}. A qualitative aspect of the present pinwheels is that their degeneracy is lifted as soon as $x\!\neq\!1$ since then the corresponding tunneling amplitude becomes finite, see Table~\ref{tab:qdm} of App.~\ref{app:nnvb}. An ED study of cluster (\ref{eq:AAsquare}) confirms this qualitative change at $x\!=\!1$. The same study reveals that the pinwheel states $|{\sf PW}_{\pm}\rangle$ are combinations of the localized bound singlets of Sec.~\ref{sec:LocalSinglets}, which provides an alternative and intuitive picture for their condensation. Finally we remark that the VBC state of Fig.~\ref{fig:pinwheels}(b) has been proposed earlier on the basis of a  large-N theory~\cite{SquaKagomelargeN}.

\begin{figure*}
\includegraphics[width=0.85\textwidth]{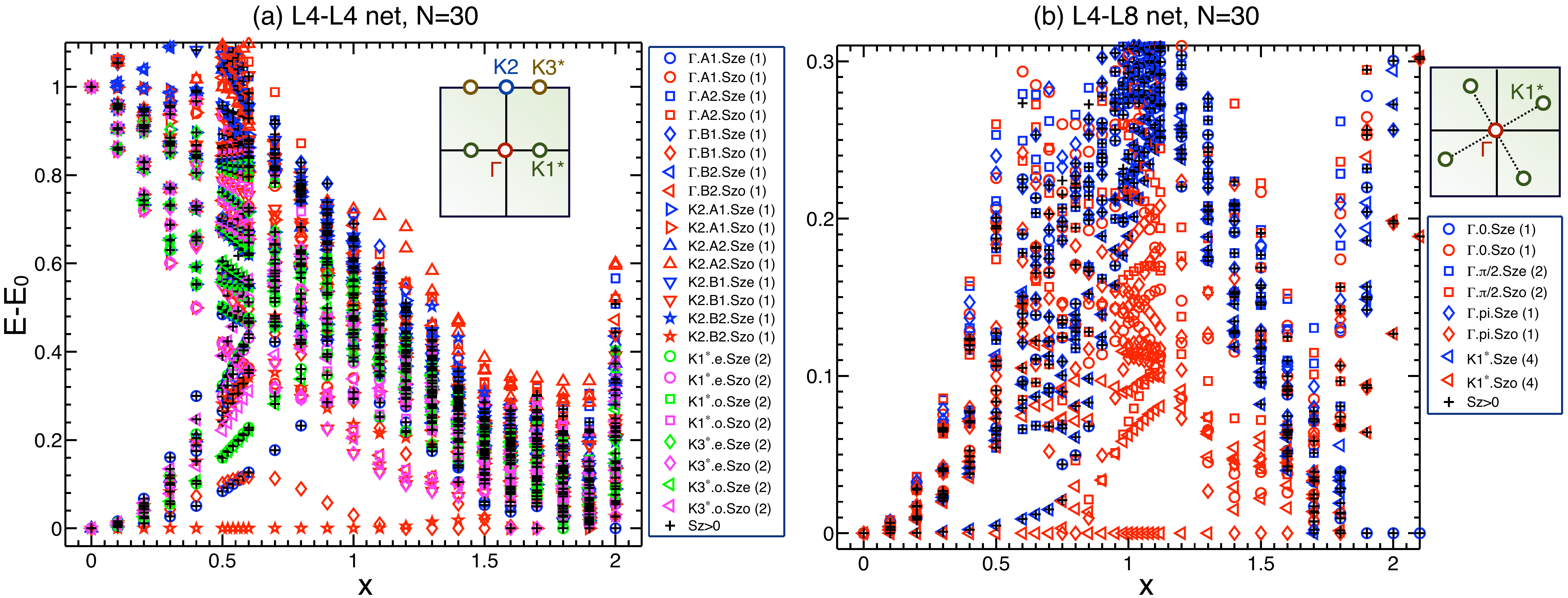}
\caption{Low-energy spectra of the 30-site L4-L4 (a) and the 30-site L4-L8 (b) clusters. Label conventions are as in Fig.~\ref{fig:spectra}.}
\label{fig:spectra30sites}
\end{figure*}

On the ED side, the situation in L4-L4 and L4-L8 is not conclusive due to the small number of unit cells in our largest clusters available ($N=20$, $30$ in L4-L4, and $N=24$, $30$ in L4-L8, see Fig.~\ref{fig:clusters} in App.~\ref{app:30sites}). One complication arises e.g. in the 20-site L4-L4, where both the effective $J_2$ (${\sf A}\!=\!0$ manifolds) and $J_{\sf BB}$ (${\sf A}\!=\!1$ manifolds) are much larger than the ones expected for the infinite system [Eq.~(\ref{eq:J2}) and below Eq.~(\ref{eq:AA1})], due to artificial contributions from the periodic boundary conditions. Similar artificial effects exist in the other clusters as well. With this in mind, the available data give the following pieces of evidence. The symmetries of the lowest four singlets highlighted by the (red) oval in Fig.~\ref{fig:spectra}(b) suggest a d-wave staggering along the horizontal or the vertical direction~\cite{footnote60}, which is not compatible with the VBC state of Fig.~\ref{fig:pinwheels}(a). Furthermore, the point group symmetries show that the state breaks the four-fold axis symmetry around AA-squares as well as the horizontal and vertical reflections. A striking aspect is that the magnetization process in the ``X'' region proceeds in steps of $\Delta S_z\!>\!1$ at low fields, which is a typical signature of spin nematic phases~\cite{shannon}. While the 30-site spectra offer an indication for this scenario (see App.~\ref{app:30sites}), another possibility is that the large magnetization steps are related to the quadrupolar-quadrupolar binding terms of Eq.~(\ref{eq:AA1}). 

The phase ``Yb'' in L4-L8 shows also $\Delta S_z\!>\!1$ magnetization steps. Since this lattice shares the same basic AA-unit with L4-L4, it is plausible that the physical origin of this feature is common in the two lattices.

\section{Summary}
We have studied three highly frustrated antiferromagnets which are built from corner-sharing triangles and feature, unlike the kagome, more than one set of inequivalent short resonance loops. We have resolved two opposite limits and uncovered very interesting physics in the middle. Away from intermediate coupling, these magnets show a number of generic features, including Lieb ferrimagnetism, frustrating $J_1$-$J_2$ physics with $J_2\!\gg\!J_1$, a sequence of fractional magnetization plateau, as well as localized magnetic and non-magnetic singlet modes. The large binding energy of the latter and their localized nature provide a clear explanation for the large number of low-lying singlets at intermediate coupling, and offer a natural connection to the short-range valence bond physics in this regime. 

Generically, the competition between inequivalent tunneling processes at intermediate coupling stabilizes valence bond crystals made of resonating building blocks. Our exact diagonalization results for the L6-L4 lattice are fully consistent with this nearest neighbor valence bond picture, showing a valence bond crystal state which breaks the six-fold rotational symmetry down to two-fold. The situation for the L4-L4 and L4-L8 lattices at intermediate coupling remains unclear. The available exact diagonalization data are not consistent with the nearest neighbor valence bond predictions and rather point at a different type of symmetry breaking and possible spin-nematic physics. This suggests that quantum fluctuations beyond the nearest neighbor valence bond basis (i.e. the influence of longer range singlets) play an important role in L4-L4 and L4-L8 lattices. This regime is theoretically most challenging and warrants further theoretical attention and experimental work on appropriate compounds.

\acknowledgments
We are grateful to A. M. L\"auchli for many useful discussions and collaborations on related work. We also thank A. Ralko and J. Richter for valuable discussions and comments. IR was supported by the Deutsche Forschungsgemeinschaft (DFG) under the Emmy-Noether program.

\appendix

\section{30-site L4-L4 and L4-L8 clusters}\label{app:30sites}
\begin{figure*}[!t]
\includegraphics[width=0.9\textwidth]{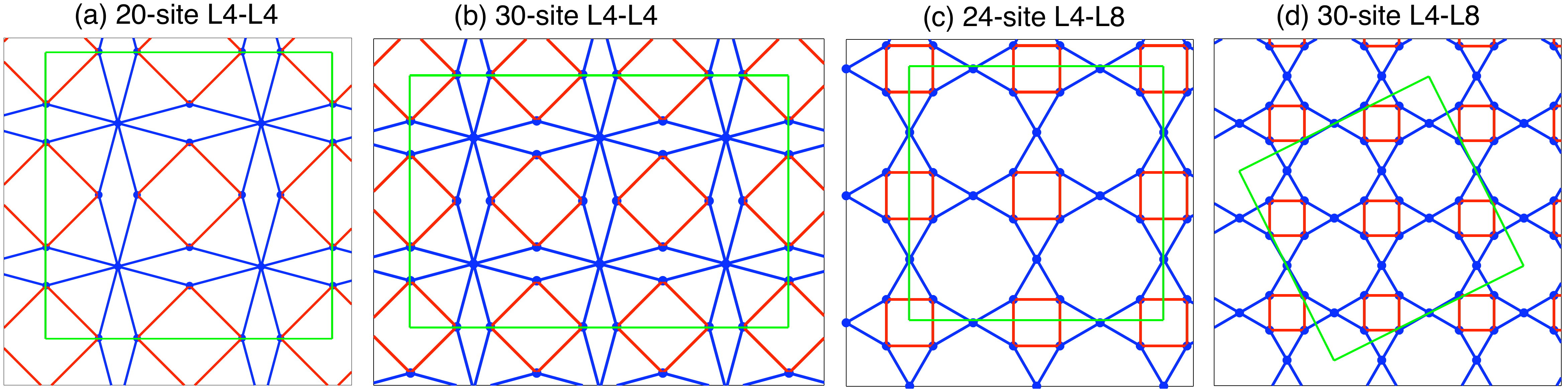}
\caption{Finite L4-L4 and L4-L8 clusters studied here (enclosed by the green line). The 32-site L6-L4 cluster is shown in Fig.~\ref{fig:LRDimer}.}
\label{fig:clusters}
\end{figure*}

Figures~\ref{fig:spectra30sites} show the low-energy spectra of the 30-site L4-L4 and L4-L8 clusters, which in turn are shown in Figs.~\ref{fig:clusters}(b) and (d). As already seen in Figs.~\ref{fig:PhaseDiagram}(b-c), these clusters give almost identical results with the smaller (symmetric) clusters for the boundaries between different field-induced phases. However, the lack of some of the symmetries of the infinite lattice becomes an issue in the intermediate coupling region, which is governed by strong competing effects. This is partly reflected in the ground state energy per site at e.g. $x\!=\!1$, which is slightly higher than the ones we find in the smaller (symmetric) clusters. 

Despite this, the low-energy spectra show many similarities with the ones in the smaller clusters. For example, the ground state level crossings suggest again one intermediate phase ``X'' in L4-L4 and two intermediate phases ``Ya'' and ``Yb'' in L4-L8. Furthermore, the two lowest singlets in phase ``X'' carry momenta $k\!=\!(0,0)$ and $(0,\pi)$, which suggests a staggering along the vertical direction of the cluster (a staggering along the horizontal direction is not commensurate with this cluster, see Fig.~\ref{fig:clusters}(b)). Finally, both ``X'' and ``Yb'' show magnetization steps with $\Delta S_z\!>\!1$.   

A quick inspection of the 30-site L4-L4 spectrum reveals that the $k\!=\!(0,0)$ singlet which crosses the ground state around $x\!\sim\!1.1$ is actually connected to a singlet at $x\!=\!0$ which minimizes fully all $K_{\sf x}$ effective terms of Sec.~\ref{sec:EffTheories}. This suggests that the intermediate phase ``X'' is stabilized by $K_{\sf x}$, in analogy with the intermediate phase of the Cairo pentagonal AFM~\cite{pentag1}. Since this lattice shares the same basic AA-unit with L4-L8, it is plausible that a large-$K_x$ mechanism may also be operative in the ``Yb'' region of L4-L8.  Exploring this scenario then requires a detailed study of the $K_x$-model in the checkerboard lattice of the B-spins.    


\section{Plateaux boundaries at small $x$}\label{app:PlateauBoundaries}
The ED data shown in Fig.~\ref{fig:PhaseDiagram} show that the boundaries between subsequent  magnetization plateaux scale linearly with $x$ with slope one. The origin of this behavior can be explained by considering the energy cost to excite a single AA-plaquette from $S_{\sf AA}\!=\!\sigma-1$ to $S_{\sf AA}\!=\!\sigma$, with $\sigma\!\ge\!2$. Since the B-spins are fully polarized we may replace (by virtue of Wigner-Eckart's theorem) any local B-spin operator with $\vec{S}_{\sf B, tot}/N_B$, where $S_{\sf B, tot}\!=\!N_{\sf B}/2$ is the total spin of the B-sublattice. Then the coupling between the A-spins of a given AA-plaquette and the surrounding B-spins reads (in units of $J_{\sf AA}$):
\be
\mc{H}_{\sf eff,\sigma}=E_{0,\sigma}+\frac{2x}{N_B} \vec{S}_{\sf AA}\cdot \vec{S}_{\sf B, tot}+\mc{O}(x^2)~,
\ee  
where $E_{0,\sigma}$ is the energy cost at $x\!=\!0$ (measured from the ground state). Solving $\mc{H}_{\sf eff,\sigma}$ analytically gives the energy of the maximum spin state, $S\!=\!\frac{N_{\sf B}}{2}\!+\!\sigma$, in a magnetic field $H$: $E_{\sigma}\!=\!E_{0,\sigma}\!+\!\sigma x \!-\! (\frac{N_{\sf B}}{2}\!+\!\sigma) H$. Setting $E_{\sigma}\!=\!E_{\sigma-1}$, gives the boundary 
\be
H_{c}^{(\sigma-1,\sigma)} = E_{0,\sigma}-E_{0,\sigma-1}+x + \mc{O}(x^2)~,
\ee
in agreement with the exact diagonalization results of Fig.~\ref{fig:PhaseDiagram}.

\section{Strong-coupling expansion in AA-triplet manifolds}\label{app:pert} 
Let us denote by $P_1$ the projection in the $A\!=\!1$ manifold (at $x\!=\!0$) of cluster (\ref{eq:AAsquare}), and by $\mc{R}_1\!=\!\frac{1-P_1}{E_1-\mc{H}}$ the resolvent operator. The second order terms where two B-sites $B_k$ and $B_l$ interact via excitations around the $A\!=\!1$ manifold by their couplings to the sites $A_i$ and $A_j$ respectively, are given by a sum of all possible terms of the type (in units of $x^2$): 
\bea\label{eq:Oijkl}
\mc{O}_{ik,jl}\!=\!P_1(\vec{A}_i\!\cdot\!\vec{B}_k)\mc{R}(\vec{A}_j\!\cdot\!\vec{B}_l)P_1\!=\!\mc{A}_{ij}^{\alpha\beta}B_k^\alpha B_l^\beta,~~~
\eea
where $\mc{A}_{ij}^{\alpha\beta}\!\equiv\!P_1 A_i^\alpha \mc{R} A_j^\beta P_1$. By Wigner-Eckart's theorem
\bea\label{eq:Aij}
&&\!\!\mc{A}_{ij}^{\alpha\beta}\!=\!\frac{1}{3}\text{Tr}\mc{A}_{ij}\delta^{\alpha\beta}
\!+\!\frac{1}{2}(\mc{A}_{ij}^{\alpha\beta}\!+\!\mc{A}_{ij}^{\beta\alpha}\!-\!\frac{2}{3}\text{Tr}\mc{A}_{ij}\delta^{\alpha\beta})\nonumber\\
\!&+&\!\frac{1}{2}(\mc{A}_{ij}^{\alpha\beta}\!-\!\mc{A}_{ij}^{\beta\alpha})
\!\equiv\! \mu_{ij}\delta^{\alpha\beta}{\bf 1}_A\!+\!\frac{1}{2}\nu^+_{ij}Q_A^{\alpha\beta}\!+\!\frac{i}{2} \nu^-_{ij}\epsilon^{\alpha\beta\gamma} A^\gamma,~~~~~~~
\eea 
where ${\bf 1}_A$ is the identity operator in the A-spin space, $\vec{A}\!=\!\sum_{i}\!\vec{A}_i$, and 
$Q_A^{\alpha\beta}\!=\!A^\alpha A^\beta\!+\!A^\beta A^\alpha\!-\!\frac{4}{3} \delta^{\alpha\beta}$  
is the quadrupolar operator~\cite{penc}. The coefficients $\mu_{ij}$ and $\nu^{\pm}_{ij}$ can be found e.g. by 
\bea
\mu_{ij}\!=\!\frac{1}{3}\!\sum_{\alpha=x,y,z}\!\!\langle t_1|\mc{A}_{ij}^{\alpha\alpha}|t_1\rangle,~
\nu^\pm_{ij}\!=\!\pm\sqrt{2} \langle t_1|\mc{A}_{ij}^{xz}\!\pm\!\mc{A}_{ij}^{zx}|t_0\rangle\nonumber~,
\eea
where $t_{\pm1,0}$ denote the three components of the AA-triplet, with their relative phases fixed to be the standard ones (a different choice of phases gives different signs for some coupling parameters). From these relations it also follows that $\mu_{ij}\!=\!\mu_{ji}$ and $\nu_{ij}^\pm\!=\!\nu_{ji}^\pm$.  

Replacing (\ref{eq:Aij}) into (\ref{eq:Oijkl}) yields
\bea\label{eq:Oijkl2}
\mc{O}_{ik,jl}\!=\!\mu_{ij}\vec{B}_k\!\cdot\!\vec{B}_l 
\!+\!\frac{\nu^+_{ij}}{16} Q_{A}^{\alpha\beta}Q_{kl}^{\alpha\beta}
\!+\!\frac{i\nu^-_{ij}}{2}\vec{A}\!\cdot\!\vec{B}_k\!\times\!\vec{B}_l,~~~~~
\eea
where $Q_{kl}^{\alpha\beta}\!\!=\!4(B_k^\alpha B_l^\beta\!+\!B_k^\beta B_l^\alpha\!-\!\frac{2}{3}\vec{B}_k\!\cdot\!\vec{B}_l\delta^{\alpha\beta})$~\cite{footnote90}. This reduces to zero if $B_{kl}\!=\!0$ or to $Q_{kl}^{\alpha\beta}\!=\!B_{kl}^\alpha B_{kl}^\alpha\!+\!B_{kl}^\alpha B_{kl}^\alpha\!-\!\frac{4}{3}\delta^{\alpha\beta}\!=\!Q_{B_{kl}}^{\alpha\beta}$ if $B_{kl}\!=\!1$. Finally the last term of (\ref{eq:Oijkl2}) simplifies to $-\frac{\nu^-_{ij}}{2}\vec{A}\!\cdot\!\vec{B}_k$ for $k\!=\!l$, thus renormalizing the corresponding first order terms discussed in the main text. For  $k\!\neq\!l$, on the other hand, the total amplitude of the unphysical term $i\vec{A}\!\cdot\!\vec{B}_k\!\times\!\vec{B}_l$ vanishes when we add contributions from $\mc{O}_{ik,jl}$ and $\mc{O}_{jl,ik}$.

\section{NNVB basis}\label{app:nnvb}
The basic ingredients of the NNVB basis are summarized in Table~\ref{tab:nnvb}.

{\it Defect triangles} --- 
In L4-L6, the total number of triangles is $N_{\sf t}=3N/4$, and so the total number of valence bonds $N_{\sf vb}\!=\!N/2\!=\!2N_{\sf t}/3$. This means that the ratio of defect triangles $N_{\sf dt}$ to $N_{\sf t}$ is $1/3$. Similarly, for L4-L4 we find $N_{\sf vb}=5N_{\sf t}/8$ and $N_{\sf dt}/N_{\sf t}=3/8$, while for L4-L8 we have $N_{\sf vb}=3N_{\sf t}/4$ and $N_{\sf dt}/N_{\sf t}=1/4$. 

{\it Energy distribution among different bonds} --- 
Consider the average expectation value of $\langle \vec{S}_i\!\cdot\!\vec{S}_j\rangle$ over AA- and BB-bonds, defined as $e_{\sf AA}$ and $e_{\sf AB}$, respectively. We can give a rough estimate for these energy distributions for the region $x\!\simeq\!1$ by disregarding the effect of NNVB resonances, i.e. by taking the expectation value of the Heisenberg Hamiltonian in any NNVB covering. This gives 
\be\label{eq:E0}
\mc{E}^{(0)}(x)=-\frac{3}{4} \left( N_{\sf vb}^{\sf AA} + x N_{\sf vb}^{\sf AB}\right)~,
\ee
where $N_{\sf vb}^{\sf AA}$ and $N_{\sf vb}^{\sf AB}$ is the number of VBs residing on AA-bonds and AB-bonds, respectively. The contribution from the AA-bonds is shared among the total number of $N_{\sf AA}$ bonds, and so we may approximate $e_{\sf AA}$ by $e_{\sf AA}^{(0)}\!=\!-\frac{3}{4}N_{\sf vb}^{\sf AA}/N_{\sf AA}$. Similarly $e_{\sf AB}^{(0)}\!=\!-\frac{3}{4}N_{\sf vb}^{\sf AB}/N_{\sf AB}$. These estimates are provided in Table~\ref{tab:nnvb} for all lattices.  

For the particular case of L4-L4, both AA- and AB-loops are squares and so, at first sight, they should equally affect the NNVB energetics of the system, at least close to $x\!=\!1$ where the two plaquettes give the same tunneling amplitudes (see below). However, because of the large connectivity of the B-sites, $N_{\sf vb}^{\sf AB}$ is smaller than $N_{\sf vb}^{\sf AA}$, despite the fact that $N_{\sf AB}\!=\!2N_{\sf AA}$. This is reflected both in $e_{\sf AB}^{(0)}$ and $e_{\sf AA}^{(0)}$, but also in the ED values $e_{\sf AB}$ and $e_{\sf AB}$ (last column of Table~\ref{tab:nnvb}). Furthermore, the fact that $e_{\sf AA}\!-\!e_{\sf AA}^{(0)}$ is more negative than $e_{\sf AB}\!-\!e_{\sf AB}^{(0)}$ shows that resonances are more effective around AA-squares.


%
\begin{figure}[!b]
\includegraphics[width=0.38\textwidth]{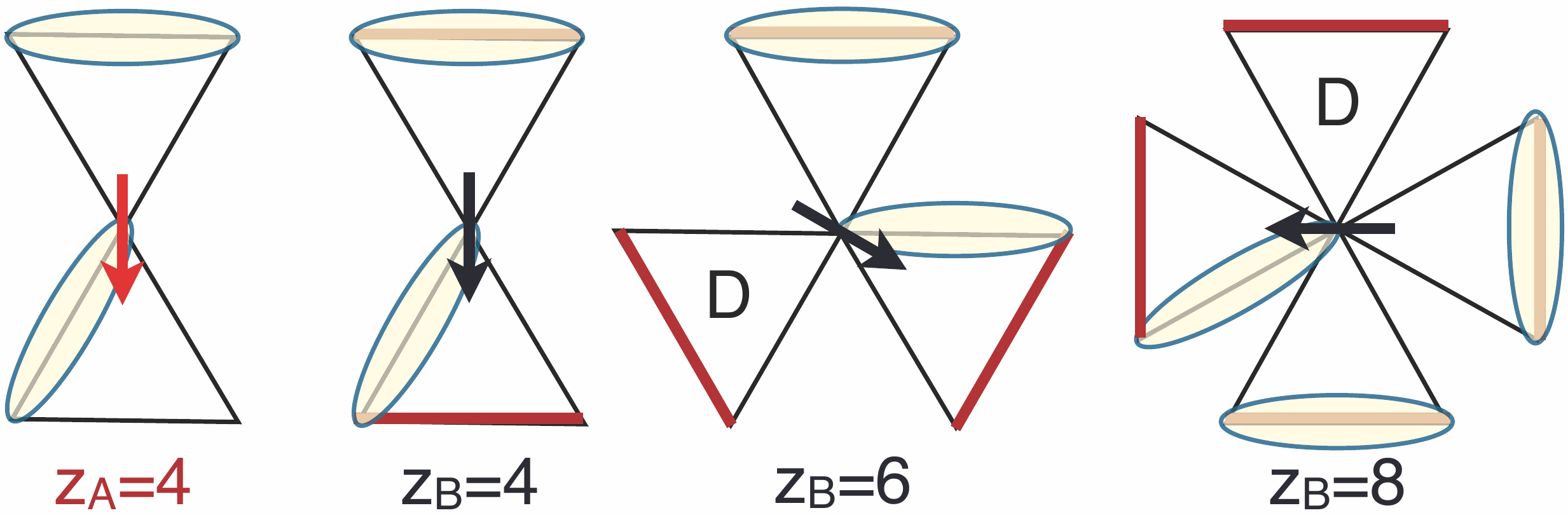} 
\caption{Arrow representation of VB coverings. The VB involving the central site must reside in one of the $z/2$ triangles sharing the site, and this is specified by an arrow with $z/2$ possible directions. ``D'' indicates defect triangles.}
\label{fig:GenArrowRepresentation}
\end{figure}

{\it Dimension of NNVB basis} --- 
To find the dimension $\mc{D}_{\sf NNVB}$ of the NNVB basis we use the following generalized arrow representation~\cite{ElserZeng93}. Consider a type-A site which, in all lattices, is shared by two triangles. The valence bond involving this site must belong to one of the two triangles, and this can be specified by introducing a (type-A) arrow with two possible directions, see Fig.~\ref{fig:GenArrowRepresentation}. Similarly, B-sites are shared by $z_{\sf B}/2$ triangles, and so the VB involving such sites can be specified by (type-B) arrows with $z_{\sf B}/2$ possible directions, namely three (L6-L4), four (L4-L4) or two (L4-L8). So in each lattice we introduce two types of arrows, one with $z_{\sf A}/2\!=\!2$ and the other with $z_{\sf B}/2$ possible directions. This gives $2^{N_{\sf A}}\left(\frac{z_{\sf B}}{2}\right)^{N_{\sf B}}$ arrow states in total. However, not all of them are faithful representations of VB coverings, since we must impose the constraint that each triangle may host at most one VB. In the arrow representation, this constraint amounts to having either no arrow pointing in or two arrows pointing in and one pointing out from each triangle. From the total $2z_{\sf B}$ arrow states per triangle, only $z_{\sf B}$ satisfy this constraint. So in each triangle we must disregard half of the arrow states. For periodic boundary conditions, this amounts to imposing $N_{\sf t}\!-\!1$ constraints in the counting. So we estimate
\be
\mc{D}_{\sf NNVB} \simeq 2^{N_{\sf A}} \left(\frac{z_{\sf B}}{2}\right)^{N_{\sf B}} / 2^{N_{\sf t}-1} = 2 \left(\frac{z_{\sf B}}{2}\right)^{N_{\sf B}}~,
\ee
where we used the equality $N_{\sf t}\!=\!N_{\sf A}$. 
While for kagome and for the L4-L8 the above counting procedure gives the correct number~\cite{Misguich03}, the numbers for L6-L4 and L4-L4 are only an estimate, similar to the Pauling counting~\cite{Pauling} of the spin ice states~\cite{SpinIce}.

\begin{figure}[!b]
\includegraphics[width=0.49\textwidth]{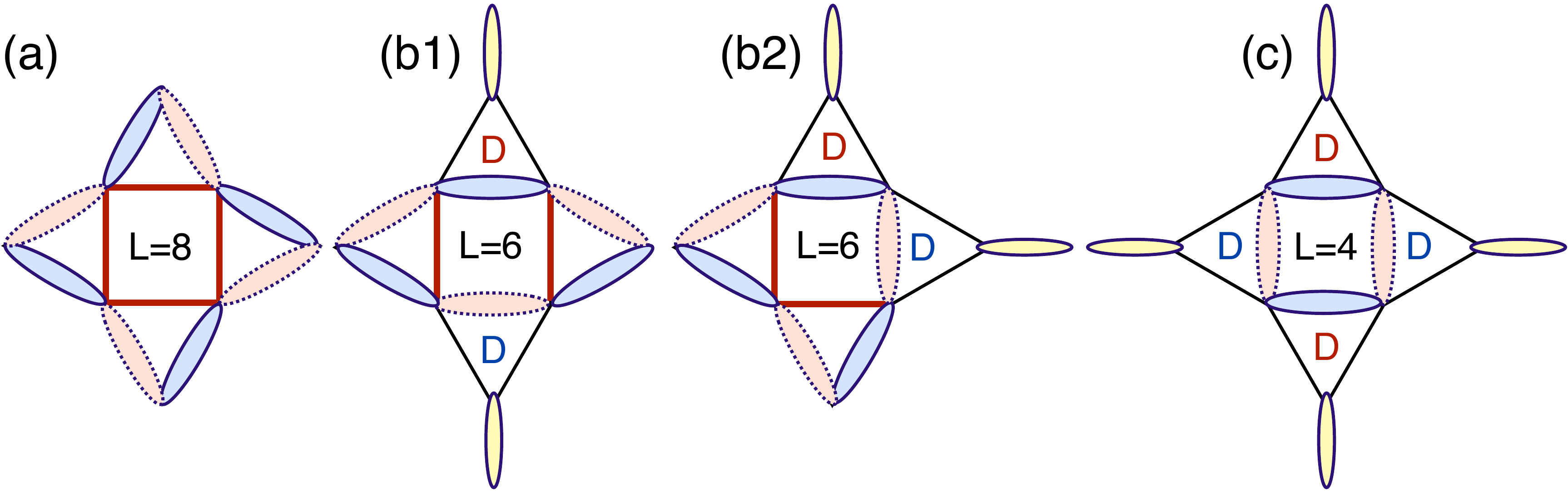} 
\caption{Finite clusters used for the evaluation of the QDM parameters for the four most local processes around AA-squares. Solid (blue) and dashed (red) ovals indicate the two dimer states $|1_p\rangle$ and $|2_p\rangle$ involved in each process, while the (red and blue) letters ``D'' denote the defect triangles in each state.}
\label{fig:RVBSquare}
\end{figure}

{\it Variational QDM parameters} --- 
Quite generally, the QDM Hamiltonian is a sum over local processes $p$, each involving two dimer states $|1_p\rangle$ and $|2_p\rangle$, as  
\bea
\mc{H}_{\sf QDM} \!&=&\! \sum_p t_p \left( |1_p\rangle\langle 2_p| + |2_p\rangle\langle 1_p| \right) \nonumber\\
&&~
+ \left( V_{1,p} |1_p\rangle\langle 1_p| +  V_{2,p} |2_p\rangle\langle 2_p| \right)~,
\eea
where $t_p$ and $V_{j,p}$ ($j\!=\!1,2$), stand for the tunneling amplitude and the polarization energies, respectively. These parameters can be found by variationally projecting the Heisenberg Hamiltonian into the NNVB basis. Owing to the non-orthogonality of the latter, this projection is equivalent to~\cite{Sutherland,ZengElser95, Mambrini2010ab} $\mc{H}_{\sf QDM}=\mc{O}^{-1/2}\mc{H} \mc{O}^{-1/2}$, where $\mc{O}$ is the overlap matrix. In a recent study~\cite{Mambrini2010ab}, $\mc{O}^{-1/2}$ was treated by an infinite-order linked-cluster diagrammatic expansion. As shown in Ref.~[\onlinecite{ioannis}], for the most local processes this method is actually equivalent with solving the 2$\times$2 NNVB problem on finite clusters that accommodate $|1_p\rangle$ and $|2_p\rangle$ as the only dimer coverings. For example, to find the four most local processes around AA-squares we use the clusters of Fig.~\ref{fig:RVBSquare}. Using $\omega_p\!=\!\langle1_p|2_p\rangle$, $E_{j,p}\!=\!\langle j_p|\mc{H}|j_p\rangle$, and ${\sf v}_p\!=\!\langle1_p|\mc{H}|2_p\rangle$, which can be obtained from the transition graph of $|1_p\rangle$ and $|2_p\rangle$ following standard rules~\cite{Sutherland, Misguich02, Mambrini2010ab}, and inverting the 2$\times$2 matrix $\mc{O}$ yields
\be
t_p\!=\!\frac{\omega_p}{1\!-\!\omega_p^2}\left(\frac{{\sf v_p}}{\omega_p}\!-\!E_{0,p}\right),
V_{1-2,p}\!=\! E_{0,p}\!-\!\omega_p t_p \!\pm\! \frac{\delta E_p}{2\sqrt{1\!-\!\omega_p^2}}~, 
\ee  
where we have defined $E_{0,p}\!\equiv\!\frac{E_{1,p}\!+\!E_{2,p}}{2}$ and $\delta E_p\!\equiv\!E_{1,p}\!-\!E_{2,p}$. Here, for all processes $\delta E_p\!=\!0$ and so $V_{1,p}\!=\!V_{2,p}\!\equiv\!V_p\!=\!E_{0,p}-\omega_p t_p$. The latter can actually be replaced by $V_p'\equiv -\omega_p t_p$, since~\cite{footnote40} $\sum_p E_{0,p}\!=\!\mc{E}^{(0)}$ is constant, see (\ref{eq:E0}).

\begin{table*}[t]
\ra{1.3} 
\begin{ruledtabular}
\begin{tabular}{@{}r|llllll@{}} 
plaquette & loop length $L_p$ & $\omega_p$ & ${\sf v}_p/\omega_p$ & $E_{0,p}$ & $t_p$ & $V_p'$ \\
\hline
AA-square & $4$ & $+2^{-1}$ & $-3$ & $-\frac{3}{2}$ & $-1$ & $+\frac{1}{2}$\\
                  & $6$ & $-2^{-2}$ & $-3x$ & $-\frac{3}{4}(2x+1)$ & $+\frac{1}{5}(2x-1)$ & $+\frac{1}{20}(2x-1)$\\
                  & $8$ & $+2^{-3}$ & $-3(2x-1)$ & $-3x$ & $+\frac{8}{21}(1-x)$ & $-\frac{1}{21}(1-x)$\\
\hline
AB-square & $4$ & $+2^{-1}$ & $-3x$ & $-\frac{3}{2}x$ & $-x$ & $+\frac{x}{2}$\\
                  & $6$ & $-2^{-2}$ & $-\frac{3}{2}(1+x)$ & $-\frac{3}{4}(2x+1)$ & $+\frac{1}{5}$ & $+\frac{1}{20}$\\
                  & $8$ & $+2^{-3}$ & $-3$ & $-\frac{3}{2} (1+x)$ & $-\frac{4}{21}(1-x)$ & $+\frac{1}{42}(1-x)$\\ 
\hline
AA-hexagon & $6$ & $-2^{-2}$ & $-\frac{9}{2}$ & $-\frac{9}{4}$ & $+\frac{3}{5}$ & $+\frac{3}{20}$\\ 
                    & $8$ & $+2^{-3}$ & $-\frac{3}{2}(2x+1)$ & $-\frac{3}{2}(x+1)$ & $-\frac{4}{21}x$ & $+\frac{1}{42}x$\\ 
                    & $10$ & $-2^{-4}$ & $-\frac{3}{2}(4x-1)$ & $-\frac{3}{4}(4x+1)$ & $+\frac{4}{85}(4x-3)$ & $+\frac{1}{340}(4x-3)$\\ 
                    & $12$ & $+2^{-5}$ & $-\frac{9}{2}(2x-1)$ & $-\frac{9}{2}x$ & $+\frac{48}{341}(1-x)$ & $-\frac{3}{682}(1-x)$\\ 
\hline
AB-octagon & $8$   & $+2^{-3}$ & $-6x$ & $-3x$ & $-\frac{8}{21}x$ & $+\frac{1}{21}x$\\                    
                   & $10$ & $-2^{-4}$ & $-\frac{3}{2}(3x+1)$ & $-\frac{3}{4}(4x+1)$ & $+\frac{4}{85}(2x+1)$ & $+\frac{1}{340}(2x+1)$\\       
                   & $12$ & $+2^{-5}$ & $-3(x+1)$ & $-\frac{3}{2}(2x+1)$ & $-\frac{16}{341}$ & $+\frac{1}{682}$\\ 
                   & $14$ & $-2^{-6}$ & $-\frac{3}{2}(x+3)$ & $-\frac{3}{4}(4x+3)$ & $+\frac{16}{1365}(3-2x)$ & $+\frac{1}{5460}(3-2x)$\\ 
                   & $16$ & $+2^{-7}$ & $-6$ & $-3(1+x)$ & $-\frac{128}{5461}(1-x)$ & $+\frac{1}{5461}(1-x)$
\end{tabular}
\end{ruledtabular}
\caption{QDM parameters $t_p$ and $V_p'$ (in units of $J_{\sf AA}$) for the most local processes around AA-squares (lines 1-3), AB-squares (lines 4-6), AA-hexagons (lines 7-10), and AB-octagons (lines 11-15).  $L_p$ is the length of the loop in the transition graph of $|1_p\rangle$ and $|2_p\rangle$. For the dimer orientations we follow the convention that singlets are oriented~\cite{footnote4} clockwise in each even-length loop. Here $E_{0,p}$ and ${\sf v}_p$ do not include the contributions from the outer (yellow) bonds. Including them does not affect $t_p$ and $V'_p$.}
\label{tab:qdm}
\end{table*}

The results for the most local processes around all plaquettes of our lattices are provided in Table~\ref{tab:qdm}. As expected, the tunneling magnitudes grow for decreasing loop lengths $L$. The dominant process for each given plaquette is the one involving the shortest loop length $L$ and the maximum number of defect triangles. Adopting the terminology for the kagome, we term these as the ``perfect resonance'' processes. Depending on the sign of $t$, these processes favor either the even or the odd combination of the two dimer coverings of the central plaquette. Pictorially, for AA-squares:
\be\label{eq:PRpm}
|{\sf PR}_{\pm}\rangle\!=\!\frac{1}{\sqrt{2(1\pm1/2)}}\left[
\parbox{0.28in}{\epsfig{file=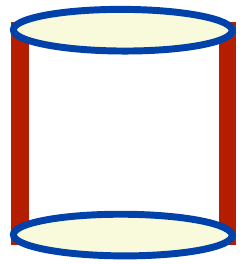,width=0.26in,clip=}}
\pm
\parbox{0.28in}{\epsfig{file=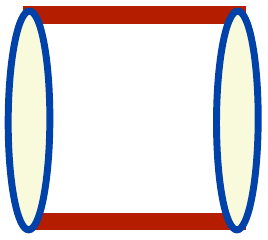,width=0.28in,clip=}}
\right]~.
\ee
According to Table~\ref{tab:qdm}, $t\!<\!0$ for squares and octagons, favoring $|{\sf PR}_+\rangle$, but for hexagons $t\!<\!0$, favoring $|{\sf PR}_-\rangle$. These states have actually a very large overlap $r$ with the exact singlet ground state of the corresponding isolated plaquette: $r\!=\!1$ for squares, $r\!=\!0.98564(5)$ for hexagons, and $r\!=\!0.94683(2)$ for octagons.

Of particular interest are the so-called ``pinwheel'' processes mentioned above, which involve the maximum possible $L$ (for a given plaquette) and no defect triangles at all. At $x\!=\!1$, the two dimer coverings involved in these processes are exact eigenstates of the cluster, and so the corresponding QDM parameters vanish. This is not true, however, for $x\!\neq\!1$. Indeed, Table~\ref{tab:qdm} shows that for all ``pinwheel'' processes, $t\propto 1-x$. Again, depending on the sign of $t$, these processes stabilize one of the two combinations of the two dimer states involved. Pictorially, for AA-squares: 
\be\label{eq:PWpm}
|{\sf PW}_{\pm}\rangle\!=\!\frac{1}{\sqrt{2(1\pm1/8)}}\left[
\parbox{0.45in}{\epsfig{file=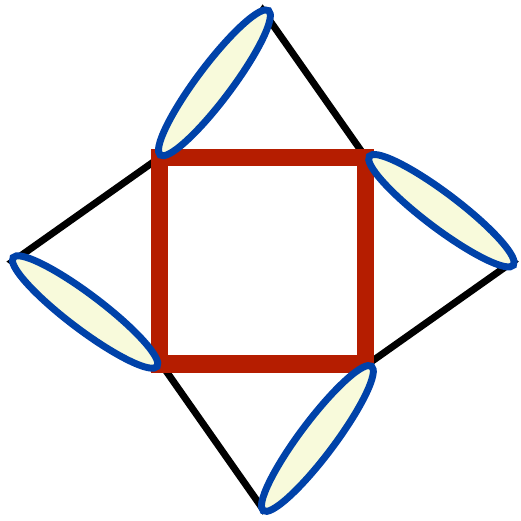,width=0.45in,clip=}}
\pm
\parbox{0.45in}{\epsfig{file=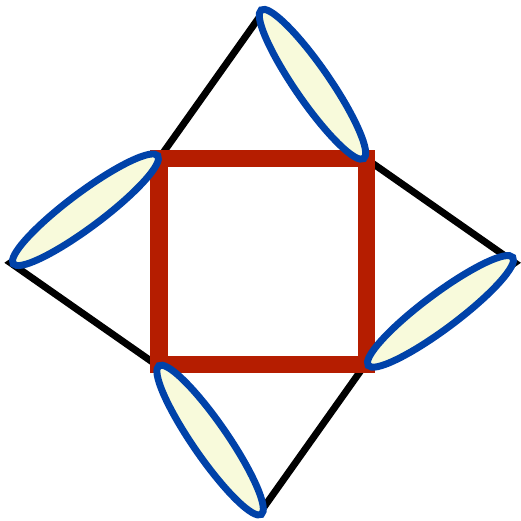,width=0.45in,clip=}}
\right]~.
\ee
For AA-plaquettes, $t\!>\!0$ (resp. $t\!<\!0$) for $x\!<\!1$ (resp. $x\!>\!1$), and thus these processes favor $|{\sf PW}_-\rangle$ for $x<1$ and $|{\sf PW}_+\rangle$ for $x\!>\!1$. At $x\!=\!1$ the two states become degenerate. This physics can be confirmed independently by an exact diagonalization of e.g. the isolated AA-square pinwheel cluster of Fig.~\ref{fig:RVBSquare}(a). 

Finally, the 6-loop resonance states around AA-squares are given by
\be\label{eq:L6pm}
|{\sf L6}_{\pm}\rangle\!=\!\frac{1}{\sqrt{2(1\mp1/4)}}\left[
\parbox{0.23in}{\epsfig{file=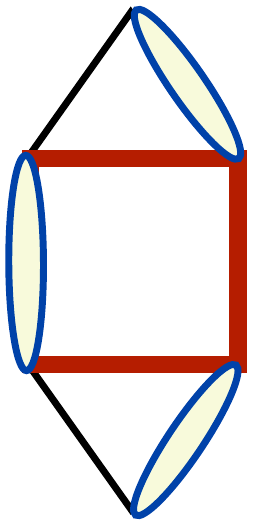,width=0.23in,clip=}}
\pm
\parbox{0.23in}{\epsfig{file=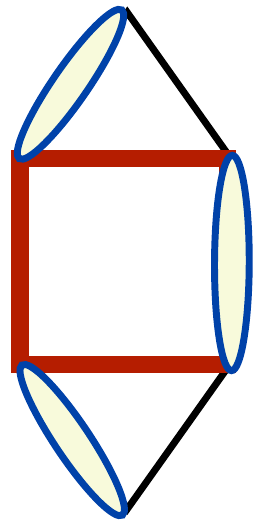,width=0.23in,clip=}}
\right]~.
\ee

\section{Classical limit}\label{app:clas} 

\begin{figure}[!b]
\includegraphics[width=0.47\textwidth]{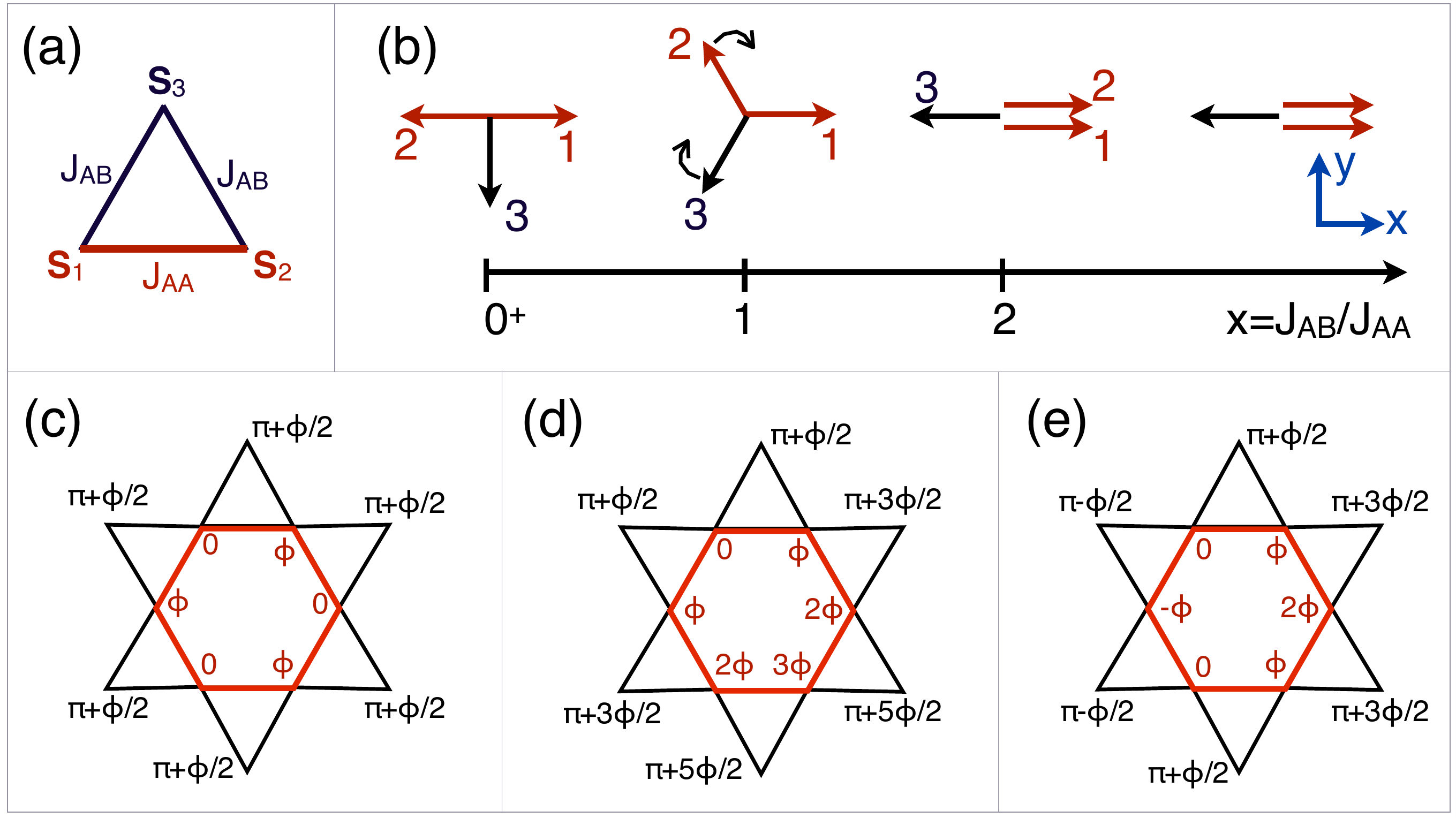} 
\caption{(a) Single-triangle contributions to the total Hamiltonian. (b) Classical phase diagram of these triangles as a function of $x\!=\!J_{\sf AB}/J_{\sf AA}$. (c-e) Some of the coplanar ground states that are possible around  the minimal AA-loops of L4-L6; directions are specified by azimuthal angles in the $xy$-plane, and $\phi$ satisfies Eq.~(\ref{eq:phi}) for the given value of $x$. In contrast to (d) and (e), in (c) all B-spins point to the same direction, allowing for weathervane modes~\cite{Chandra}.}
\label{fig:classicalminimum}
\end{figure}

All three lattices considered here have an extensive classical ground state degeneracy, which is robust in the entire region $0\!<\!x\!<\!2$. To see this, we rewrite the Hamiltonian as a sum over triangle contributions, of the type  
\be
\mc{H}_\triangle=J_{{\sf AA}} \vec{S}_1 \cdot\vec{S}_2+ J_{AB} \left( \vec{S}_1+\vec{S}_2 \right)\cdot\vec{S}_3
\ee
where $\vec{S}_{1-3}$ are the three spins on the given triangle, see Fig.~\ref{fig:classicalminimum}(a). Due to the rotational  SO(3) symmetry we are free to choose the spin plane of $\vec{S}_1$ and $\vec{S}_2$  as the $xy$-plane, with $\vec{S}_1=\vec{x}$, and $\vec{S}_2=\cos\phi_2\vec{x}+\sin\phi_2\vec{y}$. We also write $\vec{S}_3=\sin\theta_3 (\cos\phi_3\vec{x}+\sin\phi_3\vec{y}) + \cos\theta_3\vec{z}$. Minimizing with respect to $\theta_3$, $\phi_3$, and $\phi_2$ we find that the classical minimum corresponds to a three-sublattice coplanar configuration with $\theta_3=\pi/2$, $\phi_3=\pi+\frac{\phi_2}{2}$, and 
\be\label{eq:phi}
\phi_2=\left\{
\begin{array}{ll}
\cos^{-1}(\frac{x^2}{2}-1), & 0\leq x\leq 2\\
0, & x\ge 2
\end{array}
\right.
\ee
which includes the 120$^\circ$ state at the special point $x\!=\!1$, see Fig.~\ref{fig:classicalminimum}(b). 

Let us now embed these solutions on the infinite lattice, starting with a minimal AA-loop structure, shown in Fig.~\ref{fig:classicalminimum} for the L4-L6 case. We first discuss coplanar states where all B-spins point to the same direction $\vec{S}_3$. Fixing the spin $\vec{S}_3$ on a given triangle leaves two possibilities for the spins $S_{1,2}$ related to each other by a $\pi$-rotation around $\vec{S}_3$ in spin space. The spins in the next triangles are automatically fixed if all B-spins point to the same direction (there are many more solutions where the B-spins point to different directions, see below). It is easy to see that each hexagon gives two possibilities, so the discrete degeneracy of this type of solutions is $2^{N_{\text{hex}}}$. 

In analogy to the kagome~\cite{Chandra,vonDelft}, we may also build non-coplanar ground states around any hexagon where the B-spins point to the same direction $\vec{S}_3$ and by rotating the internal A-spins by the same angle. For kagome, these zero-energy local rotations are the well-known ``weathervane'' modes. 


\begin{thebibliography}{99}
\bibitem{footnote1} Since $N_{\sf B}\!<\!N_{\sf A}$ and $z_{\sf B}\!\geq\!z_{\sf A}$, spins with large coordination have large correction, which is at odds with the intuitive mean-field picture that such sites behave more classically. There are many other examples with this behavior~\cite{Moessner-dS,pentag1}.
\bibitem{footnote2} This can also seen by directly acting $\mc{H}_{K_{\sf x}}$ on a AA-square of the p-CD-VBC state. The singlet on one diagonal gives an effective FM coupling $-\frac{3}{4}K_{\sf x}\!<\!0$ along the second diagonal. In principle, the triplet portion can be saturated by a large enough $K_{\sf x}$, and the resulting $S\!=\!1$ objects would then form (via $J_2$) AFM chains which are interconnected via $J_1$. Such a network of Haldane chains (which also appears in the distorted Shastry-Sutherland magnet~\cite{Marion}) may be adiabatically connected to the p-CD-VBC phase at $K_{\sf x}\!=\!0$, but the details of such a scenario are not clear.
\bibitem{footnote3a} Note that processes involving NN B-spins do not appear in second order, similarly to the $J_1$ coupling inside the ${\sf A}\!=\!0$ manifold.
\bibitem{footnote3b} The integer $n$ can be as large as a macroscopic fraction of the number of unit cells $N_{\sf uc}$, determined by the requirement that the corresponding AA-plaquettes do not touch each other.
\bibitem{footnote55} We remark here that in (b) the three excited singlets (``$\Gamma$.0.Sze'' and ``M$^\ast$\!.A2.Sze'') are degenerate owing to an enlarged T$_d$ symmetry of the 20-site L4-L4 cluster.
\bibitem{footnote50} The low-lying singlets in L4-L8 are also discussed in [\onlinecite{SquaKagomeRichter3}] for $x\!=\!1$.
\bibitem{footnote60} Note that such lattice nematicity is also present in the collinear phase at small $x$.
\bibitem{footnote90} Compared to the definition of Ref.~[\onlinecite{penc}], here we have included an extra factor of 4 so that the resulting quadrupolar operator in the $B_{kl}\!=\!1$ sector has the same form with the one for spins $S\!=\!1$.
\bibitem{footnote40} In this sum one must include the contributions from dimers that do not participate in any resonance.
\end{thebibliography}
\end{document}